\documentstyle[aps,eqsecnum,preprint,floats,epsf,epsfig]{revtex}
\begin{document}
\def\be{\begin{eqnarray}}
\def\en{\end{eqnarray}}
\def\non{\nonumber}
\def\la{\langle}
\def\ra{\rangle}
\def\nc{N_c^{\rm eff}}
\def\vp{\varepsilon}
\def\a{{\cal A}}
\def\B{{\cal B}}
\def\c{{\cal C}}
\def\d{{\cal D}}
\def\e{{\cal E}}
\def\p{{\cal P}}
\def\t{{\cal T}}
\def\up{\uparrow}
\def\dw{\downarrow}
\def\vma{{_{V-A}}}
\def\vpa{{_{V+A}}}
\def\smp{{_{S-P}}}
\def\spp{{_{S+P}}}
\def\J{{J/\psi}}
\def\ov{\overline}
\def\Lqcd{{\Lambda_{\rm QCD}}}
\def\pr{{\sl Phys. Rev.}~}
\def\prl{{\sl Phys. Rev. Lett.}~}
\def\pl{{\sl Phys. Lett.}~}
\def\np{{\sl Nucl. Phys.}~}
\def\zp{{\sl Z. Phys.}~}
\def\lsim{ {\ \lower-1.2pt\vbox{\hbox{\rlap{$<$}\lower5pt\vbox{\hbox{$\sim$}
}}}\ } }
\def\gsim{ {\ \lower-1.2pt\vbox{\hbox{\rlap{$>$}\lower5pt\vbox{\hbox{$\sim$}
}}}\ } }

\font\el=cmbx10 scaled \magstep2{\obeylines\hfill July, 2003}

\vskip 1.5 cm

\centerline{\large\bf Hadronic $B$ Decays Involving Even Parity
Charmed Mesons}
\bigskip
\centerline{\bf Hai-Yang Cheng}
\medskip
\centerline{Institute of Physics, Academia Sinica}
\centerline{Taipei, Taiwan 115, Republic of China}
\medskip

\bigskip
\bigskip
\centerline{\bf Abstract}
\bigskip
{\small Hadronic $B$ decays containing an even-parity charmed
meson in the final state are studied. Specifically we focus on the
Cabibbo-allowed decays $\ov B\to D^{**}\pi(\rho),~D^{**}\bar
D_s^{(*)},~\bar D^{**}_sD^{(*)}$ and $\ov B_s\to
D_s^{**}\pi(\rho)$, where $D^{**}$ denotes generically a $p$-wave
charmed meson. The $B\to D^{**}$ transition form factors are
studied in the improved version of the Isgur-Scora-Grinstein-Wise
quark model. We apply heavy quark effective theory and chiral
symmetry to study the strong decays of $p$-wave charmed mesons and
determine the magnitude of the $D_1^{1/2}\!-\!D_1^{3/2}$ mixing
angle (the superscript standing for the total angular momentum of
the light quark). Except the decay to $D_1(2427)^0\pi^-$ the
predictions of $\B(B^-\to D^{**0}\pi^-)$ agree with experiment.
The sign of the $D_1^{1/2}\!-\!D_1^{3/2}$ mixing angle is found to
be positive in order to avoid a severe suppression on the
production of $D_1(2427)^0\pi^-$. The interference between
color-allowed and color-suppressed tree amplitudes is expected to
be destructive in the decay $B^-\to D_1(2427)^0\pi^-$. Hence, an
observation of the ratio $D_1(2427)^0\pi^-/D_1(2427)^+\pi^-$ can
be used to test the relative signs of various form factors as
implied by heavy quark symmetry. Although the predicted $B^-\to
D_1(2420)^0\rho^-$ at the level of $3\times 10^{-3}$ exceeds the
present upper limit, it leads to the ratio
$D_1(2420)\rho^-/D_1(2420)\pi^-\approx 2.6$ as expected from the
factorization approach and from the ratio $f_\rho/f_\pi\approx
1.6$\,. Therefore, it is crucial to have a measurement of this
mode to test the factorization hypothesis. For $\ov B\to \ov
D_s^{**}D$ decays, it is expected that $\ov D_{s0}^*D\gsim \ov
D_{s1}D$ as the decay constants of the multiplet
$(D_{s0}^*,D_{s1})$ become identical in the heavy quark limit. The
experimental observation of less abundant production of $\ov
D_{s0}^*D$ than $\ov D_{s1}D$ is thus a surprise. What is the
cause for the discrepancy between theory and experiment remains
unclear. Meanwhile, it is also important to measure the $B$ decay
into $\ov D_{s1}(2536)D^{(*)}$ to see if it is suppressed relative
to $\ov D_{s1}(2460)D^{(*)}$ to test the heavy quark symmetry
relation $f_{D_{s1}(2536)}\ll f_{D_{s1}(2460)}$. Under the
factorization hypothesis, the production of $\ov D_{s2}^*D$ is
prohibited as the tensor meson cannot be produced from the $V-A$
current. Nevertheless, it can be induced via final-state
interactions or nonfactorizable contributions and hence an
observation of $\ov B\to\ov D_{s2}^*D^{(*)}$ could imply the
importance of final-state rescattering effects.

}

\pagebreak

\section{Introduction}
The interest in even-parity charmed mesons is revived by a recent
discovery of a new narrow resonance by BaBar \cite{BaBar}. This
state which can be identified with $J^P=0^+$ is lighter than most
theoretical predictions for a $0^+$ $c\bar s$ state \cite{QM}.
Moreover, a renewed lattice calculation \cite{Bali} yields a
larger mass than what is observed. This unexpected and surprising
disparity between theory and experiment has sparked a flurry of
many theory papers. For example, it has been advocated that this
new state is a four-quark bound state
\cite{CH,Nussinov},\footnote{The low-lying non-charm scalar mesons
in the conventional $q\bar q'$ states are predicted by the quark
potential model to lie in between 1 and 2 GeV, corresponding to
the nonet states $f_0(1370)$, $a_0(1450)$, $K_0^*(1430)$ and
$f_0(1500)/f_0(1710)$. The light scalar nonet formed by
$\sigma(600)$, $\kappa(800)$, $f_0(980)$ and $a_0(980)$ can be
identified primarily as four-quark states \cite{Jaffe}. It has
been argued \cite{Close} that a strong attraction between
$(qq)_{\bf 3^*}$ and $(\bar q\bar q)_{\bf 3}$ \cite{Jaffe,Alford},
where ${\bf 3^*}$ and {\bf 3} here refer to color, and the absence
of the orbital angular momentum barrier in the $s$-wave 4-quark
state, may explain why the scalar nonet formed by four-quark bound
states is lighter than the conventional $q\bar q$ nonet. By the
same token, it is likely that a scalar $cn\bar n\bar s$ 4-quark
state, where $n = u,\ d$, will be lighter than the $0^+$ $p$-wave
$c\bar s$ state, where a typical potential model prediction gives
2487~MeV~\cite{QM}. It has been suggested in \cite{CH} to search
for exotic 4-quark $cqq\bar q$ charmed meson production in $B$
decays. Particularly noteworthy are resonances in the doubly
charged $D_s^+\pi^+$ ($D^+K^+$), and wrong pairing $D^+K^-$
channels. However, contrary to the case of scalar resonances, the
$1^+$ $D_s(2460)$ state is unlikely a four-quark state as it is
heavier than the axial-vector meson formed by $c\bar s$. A
non-observation of a heavier and broad $0^+$ $c\bar s$ state will
not support the four-quark interpretation of $D_s(2317)$.} or a
$DK$ molecular \cite{Barnes} or even a $D\pi$ atom \cite{Szc}.  On
the contrary, it has been put forward that, based on heavy quark
effective theory and chiral perturbation theory, the newly
observed $D_s(2317)$ is the $0^+$ $c\bar s$ state and that there
is a $1^+$ chiral partner with the same mass splitting with
respect to the $1^-$ state as that between the $0^+$ and $0^-$
states \cite{Bardeen,Nowak}. The existence of a new narrow
resonance with a mass near 2.46 GeV which can be identified with
$1^+$ state was first hinted by BaBar and has been observed and
established by CLEO \cite{CLEO} and Belle \cite{BelleD}.

Although the $D_{s0}^*(2317)$ and $D_{s1}(2460)$ states were
discovered in charm fragmentation of $e^+e^-\to c\bar c$, it will
be much more difficult to measure the counterpart of
$D_{s0}^*(2317)$ and $D_{s1}(2460)$ in the non-strange charm
sector, namely $D_0^*$ and $D_1$, owing to their large widths.
Indeed, the broad $D_0^*$ and $D_1$ resonances were explored by
Belle \cite{BelleD} in charged $B$ to $D^+\pi^-\pi^-$ and
$D^{*+}\pi^-\pi^-$ decays (see Table \ref{tab:mass}). The study of
even-parity charmed meson production in $B$ decays, which is the
main object of this paper, also provides an opportunity to test
heavy quark effective theory.

This work is organized as follows. The masses and widths of
$p$-wave charmed mesons are summarized in Sec. II. In order to
determine the mixing angle of the axial-vector charmed mesons, we
apply heavy quark effective theory and chiral symmetry to study
their strong decays. The decay constants of $p$-wave charmed
mesons and $B\to D^{**}$ form factors are studied in Sec. III
within the Isgur-Scora-Grinstein-Wise quark model. The production
of $p$-wave charmed mesons in $B$ decays is studied in detail in
Sec. IV. Conclusions are presented in Sec. V.

\begin{table}[ht]
\caption{The masses and decay widths of even-parity charmed
mesons. We follow the naming scheme of Particle Data Group [15] to
add a superscript ``*" to the states if the spin-parity is in the
``normal" sense, $J^P=0^+,1^-,2^+,\cdots$. The four $p$-wave
charmed meson states are thus denoted by $D_0^*,D_1,D'_1$ and
$D_2^*$. In the heavy quark limit, $D_1$ has $j=1/2$ and $D'_1$
has $j=3/2$ with $j$ being the total angular momentum of the light
degrees of freedom.
 } \label{tab:mass}
\begin{center}
\begin{tabular}{l l l l }
State & Mass (MeV) & Width (MeV) & Reference \\
\hline
 $D^*_0(2308)^0$ & $2308\pm17\pm15\pm28$ & $276\pm21\pm18\pm60$ & \cite{BelleD} \\
 $D_1(2427)^0$ & $2427\pm26\pm20\pm15$ & $384^{+107}_{-~75}\pm24\pm70$ & \cite{BelleD} \\
 $D'_1(2420)^0$ & $2422.2\pm1.8$ & $18.9^{+4.6}_{-3.5}$ & \cite{PDG} \\
 & $2421.4\pm 1.5\pm0.4\pm0.8$ & $23.7\pm 2.7\pm0.2\pm
 4.0$ & \cite{BelleD} \\
 $D'_1(2420)^\pm$ & $2427\pm5$ & $28\pm8$ & \cite{PDG} \\
 $D_2^*(2460)^0$ & $2458.9\pm 2.0$ & $23\pm 5$ & \cite{PDG} \\
 & $2461.6\pm2.1\pm0.5\pm3.3$ & $45.6\pm4.4\pm6.5\pm1.6$ &  \cite{BelleD} \\
 $D_2^*(2460)^\pm$ & $2459\pm 4$ & $25^{+8}_{-7}$ & \cite{PDG} \\
 \hline
 $D_{s0}^*(2317)$ & $2317.3\pm0.4$ & $<7$ & \cite{BaBar,CLEO,BelleD} \\
 $D_{s1}(2460)$ & $2458.9\pm1.2$ & $<7$ & \cite{CLEO,BelleD} \\
 $D'_{s1}(2536)$ & $2535.35\pm0.34\pm0.5$ & $<2.3$ & \cite{PDG} \\
 $D_{s2}^*(2573)$ & $2572.4\pm1.5$ & $15^{+5}_{-4}$ & \cite{PDG} \\
\end{tabular}
\end{center}
\end{table}

\section{Mass spectrum and decay width}
In the quark model, the even-parity mesons are conventionally
classified according to the quantum numbers $J,L,S$: the scalar
and tensor mesons correspond to $^{2S+1}L_J=\,^3P_0$ and $^3P_2$,
respectively, and there exit two different axial-vector meson
states, namely, $^1P_1$ and $^3P_1$ which can undergo mixing if
the two constituent quarks do not have the same masses. For heavy
mesons, the heavy quark spin $S_Q$ decouples from the other
degrees of freedom in the heavy quark limit, so that $S_Q$ and the
total angular momentum of the light quark $j$ are separately good
quantum numbers.  The total angular momentum $J$ of the meson is
given by $\vec J=\vec j+\vec S_Q$ with $\vec S=\vec s+\vec S_Q$
being the total spin angular momentum. Consequently, it is more
natural to use $L^{j}_J=P_2^{3/2},P_1^{3/2},P_1^{1/2}$ and
$P_0^{1/2}$ to classify the first excited heavy meson states where
$L$ here is the orbital angular momentum of the light quark. It is
obvious that the first and last of these states are $^3P_2$ and
$^3P_0$, while \cite{IW}
 \be \label{PjJ}
 |P_1^{3/2}\ra=\sqrt{2\over 3}\,|\,^1P_1\ra+\sqrt{1\over 3}\,|\,^3P_1\ra,
 \qquad  |P_1^{1/2}\ra=-\sqrt{1\over 3}\,|\,^1P_1\ra+\sqrt{2\over
 3}\,|\,^3P_1\ra.
 \en
In the heavy quark limit, the physical eigenstates with $J^P=1^+$
are $P_1^{3/2}$ and $P_1^{1/2}$ rather than $^3P_1$ and $^1P_1$.

The masses and decay widths of even-parity (or
$p$-wave)\footnote{If the even-parity mesons are the bound states
of four quarks, they are in an orbital $s$-wave. In this case, one
uses $J^P=0^+$ rather $^3P_0$ to denote scalar mesons, for
example.} charmed mesons $D_J^*$ and $D_{sJ}^*$ are summarized in
Table \ref{tab:mass}. We shall use $1^+$ and $1'^+$ or $D_1$ and
$D'_1$ to distinguish between two different physical axial-vector
charmed meson states. As we shall see below, the physical $1^+$
state is primarily $P_1^{1/2}$, while $1'^+$ is predominately
$P_1^{3/2}$. A similar broad $D_1$ state not listed in Table I was
reported by CLEO \cite{Galik} with $M=2461^{+42}_{-35}$ MeV and
$\Gamma=290^{+104}_{-~83}$ MeV. For the known narrow resonances,
the Belle's measurement of the $D_2^{*0}$ width (see Table
\ref{tab:mass}) is substantially higher than the current world
average of $23\pm5$ MeV \cite{PDG}.

In the heavy quark limit, the states within the chiral doublets
$(0^+,1^+)$ with $j=1/2$ and $(1'^+,2^+)$ with $j=3/2$ are
degenerate. After spontaneous chiral symmetry breaking, $0^+$
states acquire masses while $0^-$ states become massless Goldstone
bosons. As shown in \cite{Bardeen}, the fine splitting between
$0^+$ and $0^-$ is proportional to the constituent quark mass. The
hyperfine mass splittings of the four $p$-wave charmed meson
states arise from spin-orbit interactions and tensor-force
interaction [see Eq. (\ref{massoperator}) below], while the
spin-spin interaction is solely responsible for the hyperfine
splitting within the multiplet $(0^-,1^-)$.

From Table I and the given masses of pseudoscalar and vector
charmed mesons in the PDG \cite{PDG}, it is found empirically that
the hyperfine splitting within the chiral multiplets $(0^+,1^+)$,
$(1'^+,2^+)$ and $(0^-,1^-)$ are independent of the flavor of the
light quark:
 \be
 && m(D_2^*)-m(D'_1)\approx m(D_{s2}^*)-m(D'_{s1})\approx 37\,{\rm
 MeV}, \non \\
 && m(D_{s1})-m(D_{s0}^*)\approx m(D^*_s)-m(D_s)=144\,{\rm MeV},
 \non \\
 && m(D_{1})-m(D_{0}^*)\approx m(D^*)-m(D)=143\,{\rm MeV}.
 \en
However, the fine splittings
 \be \label{finespli}
&& m(D_{s0}^*)-m(D_s)\approx m(D_{s1})-m(D_s^*)\approx 350\,{\rm
MeV},  \non \\
&& m(D_{0}^*)-m(D)\approx m(D_1)-m(D^*)\approx 430\,{\rm MeV},
  \en
depend on the light quark flavor.\footnote{Likewise, considering
the spin-averaged masses of the doublets $(0^+,1^+)$ and
$(1'^+,2^+)$
 \be
 \ov m_{01}(D)\equiv {1\over 4}m(D_0^*)+{3\over 4}m(D_1), \qquad
 \ov m_{12}(D)\equiv {3\over 8}m(D'_1)+{5\over 8}m(D_2^*),
 \en
the hyperfine mass splittings
 \be
 \ov m_{12}(D)-\ov m_{01}(D)\approx 48\,{\rm MeV},\quad \ov m_{12}
 (D_s)-\ov m_{01}(D_s)\approx 132\,{\rm MeV},
 \en
also depend on the light quark flavor. Based on a quark-meson
model, the spin-weighted masses $\ov m_{01}(D)=2165\pm 50$ MeV
\cite{Deandrea98} and $\ov m_{01}(D_s)=2411\pm25$ MeV
\cite{Deandrea03} were predicted, while experimentally they are
very similar (see Table I). } Since the fine splitting between
$0^+$ and $0^-$ or $1^+$ and $1^-$ should be heavy-flavor
independent in heavy quark limit, the experimental result
(\ref{finespli}) implies that the fine splitting is light quark
mass dependent. Indeed, if the first line rather than the second
line of Eq. (\ref{finespli}) is employed as an input for the fine
splittings of non-strange charmed mesons one will predict
\cite{Bardeen}
 \be
&& M(D_0^{*\pm})=2217\,{\rm MeV}, \qquad M(D_0^{*0})=2212\,{\rm
MeV}, \non \\ && M(D_1^\pm)=2358\,{\rm MeV}, \qquad
M(D_1^0)=2355\,{\rm MeV},
 \en
which are evidently smaller than what are measured by Belle
\cite{BelleD}.

It is interesting to note that the mass difference between strange
and non-strange charmed mesons is of order $100\sim 110$ MeV for
$0^-$, $1^-$, $1'^+$ and $2^+$ as expected from the quark model.
As a consequence, the experimental fact that $m(D_{s0}^*)\approx
m(D_0^*)$ and $m(D_{s1})\sim m(D_1)$ is very surprising.

In the heavy quark limit, the physical mass eigenstates $D_1$ and
$D'_1$ can be identified with $P_1^{1/2}$ and $P_1^{3/2}$,
respectively. However, beyond the heavy quark limit, there is a
mixing between $P_1^{1/2}$ and $P_1^{3/2}$, denoted by $D_1^{1/2}$
and $D_1^{3/2}$ respectively,
 \be \label{Dmixing}
 D_1(2427) &=& D_1^{1/2}\cos\theta+D_1^{3/2}\sin\theta, \non \\
 D'_1(2420) &=& -D_1^{1/2}\sin\theta+D_1^{3/2}\cos\theta.
 \en
Likewise for strange axial-vector charmed mesons
 \be \label{Dsmixing}
 D_{s1}(2460) &=& D_{s1}^{1/2}\cos\theta_s+D_{s1}^{3/2}\sin\theta_s, \non \\
 D'_{s1}(2536) &=& -D_{s1}^{1/2}\sin\theta_s+D_{s1}^{3/2}\cos\theta_s.
 \en
Since $D_1^{1/2}$ is much broader than $D_1^{3/2}$ as we shall see
shortly, the decay width of $D'_1(2420)$ is sensitive to the
mixing angle $\theta$. Our task is to determine the
$D_1^{1/2}\!-\!D_1^{3/2}$ mixing angle from the measured widths.
In contrast, the present upper limits on the widths of
$D_{s1}(2460)$ and $D'_{s1}(2536)$ do not allow us to get any
constraints on the mixing angle $\theta_s$. Hence, we will turn to
the quark potential model to extract $\theta_s$ as will be shown
below.

It is suitable and convenient to study the strong decays of heavy
mesons within the framework of heavy quark effective theory in
which heavy quark symmetry and chiral symmetry are combined
\cite{TMYan}. It is straightforward to generalize the formalism to
heavy mesons in $p$-wave excited states \cite{Falk}. The decay
$D_0^*$ undergoes a $s$-wave hadronic decay to $D\pi$, while
$D_1^{1/2}$ can decay into $D^*$ by $s$-wave and $d$-wave pion
emissions but only the former is allowed in the heavy quark limit
$m_c\to\infty$:
 \be \label{swave}
 \Gamma(D_0^*\to D\pi) = g_{D_0^*D\pi}^2{p_c\over 8\pi m^2_{D_0}},
 \qquad\quad
 \Gamma(D_1^{1/2}\to D^*\pi) = g_{D_1^{1/2}D^*\pi}^2{p_c\over 8\pi
 m^2_{D_1^{1/2}}},
 \en
where $p_c$ is the c.m. momentum of the final-state particles in
the $B$ rest frame. The tensor meson $D_2^*$ decays into $D^*$ or
$D$ via $d$-wave pion emission. In the heavy quark limit where the
total angular momentum $j$ of the light quark is conserved,
$D_1^{3/2}\to D\pi$ is prohibited by heavy quark spin symmetry.
The explicit expressions for the decay rates are \cite{Falk},
 \be \label{Dwaverate}
 \Gamma(D_1^{3/2}\to D^*\pi)={1\over 6\pi}\,{m_{D^*}\over
 m_{D_1^{3/2}}}\,{h'^2\over \Lambda_\chi^2}\,{p_c^5\over f_\pi^2},
 \non \\
 \Gamma(D_2^*\to D^*\pi)={1\over 10\pi}\,{m_{D^*}\over
 m_{D_2}}\,{h'^2\over \Lambda_\chi^2}\,{p_c^5\over f_\pi^2},
 \non \\
 \Gamma(D_2^*\to D\pi)={1\over 15\pi}\,{m_{D}\over
 m_{D_2}}\,{h'^2\over \Lambda_\chi^2}\,{p_c^5\over f_\pi^2},
 \en
where $\Lambda_\chi$ is a chiral symmetry breaking scale,
$f_\pi=132$ MeV and $h'$ is a heavy flavor independent coupling
constant. The $p_c^5$ dependence of the decay rate indicates the
$d$-wave nature of pion emission. From Eq. (\ref{Dwaverate}) we
obtain
 \be
 {\Gamma(D_2^{*0}\to D^+\pi^-)\over \Gamma(D_2^{*0}\to
 D^{*+}\pi^-)}={2\over 3}\,{m_D\over m_{D^*}}\,\left({p_c(D_2^*\to
 D\pi)\over p_c(D_2^*\to D^*\pi)}\right)^5=2.3\,,
 \en
in excellent agreement with the measured value of $2.3\pm 0.6$
\cite{PDG}.

Since the $d$-wave decay is severely phase-space suppressed, it is
evident that $D_0^*$ and $D_1$ are very broad, of order 250 MeV in
their widths, whereas $D'_1$ and $D_2^*$ are narrow with widths of
order 20 MeV.

The strong couplings appearing in Eq. (\ref{swave}) are given by
 \be
 g_{D_0^*D\pi} &=& \sqrt{m_{D_0}m_D}\,{m_{D_0}^2-m_D^2\over
 m_{D_0}}\,{h\over f_\pi}, \non \\
 g_{D_1^{1/2}D^*\pi} &=& \sqrt{m_{D_1^{1/2}}m_{D^*}}\,{m_{D_1^{1/2}}^2-m_{D^*}^2\over
 m_{D_1^{1/2}}}\,{h\over f_\pi},
 \en
with $h$ being another heavy-flavor independent coupling constant
in the effective Lagrangian \cite{Falk}. It can be extracted from
the measured width of $D_0^*(2308)$ (see Table I) to be
 \be \label{h}
 h=0.65\pm 0.12\,.
 \en
From the averaged width $29.4\pm 4.2$ MeV measured for $D_2^{*0}$
we obtain
 \be \label{h'}
 {h'\over \Lambda_\chi}=0.67\pm0.05\,{\rm GeV}^{-1}\,.
 \en
Substituting the couplings $h$ and $h'$ into Eqs. (\ref{swave})
and (\ref{Dwaverate}) leads to
 \be
 \Gamma(D_1^{3/2}\to D^*\pi)=10.5\pm 1.5\,{\rm MeV}, \qquad
 \Gamma(D_1^{1/2}\to D^*\pi)=248\pm92\,{\rm MeV},
 \en
where we have assumed that $D_1^{3/2}$ has a mass close to
$D'_1(2420)$ and $m(D_1^{1/2})\approx m(D_1(2427))$. Therefore,
$D_1^{3/2}$ is much narrower than $D_1^{1/2}$ owing to the
phase-space suppression for $d$-wave. However, the physical state
$D'_1(2420)$ can receive a $s$-wave contribution as there is a
mixing between $D_1^{1/2}$ and $D_1^{3/2}$ beyond the heavy quark
limit. The observed narrowness of $D'_1(2420)$ indicates that the
mixing angle should be small; that is, $D'_1(2420)$ should be
dominated by $D_1^{3/2}$ while $D_1(2427)$ is primarily the
$D_1^{1/2}$ state. Using
 \be
 \Gamma[D'_1(2420)]=\Gamma(D_1^{3/2}\to
 D^*\pi)\cos^2\theta+\Gamma(D_1^{1/2}\to D^*\pi)\sin^2\theta
 \en
and the averaged width $20.9\pm 2.1$ MeV for $D'_1(2420)^0$, it is
found \footnote{The $D_1^{1/2}\!-\!D_1^{3/2}$ mixing angle was
just reported to be $\theta=0.10\pm0.03\pm0.02\pm0.02\,{\rm
radians}=(5.7\pm2.4)^\circ$ by Belle through a detailed $B\to
D^*\pi\pi$ analysis \cite{BelleD}. This is consistent with the
result of Eq. (\ref{theta}).}
 \be \label{theta}
 \theta=\pm(12.1^{+6.6}_{-4.4})^\circ.
 \en
We shall see in Sec. IV that a positive mixing angle is preferred
by a study of $D_1(2427)^0\pi^-$ production in $B$ decays.

The scalar resonance $D_{s0}^*(2317)$ is below the threshold of
$DK$ and its only allowed strong decay $D_{s0}^*(2317)\to
D_s\pi^0$ is isospin violating. Therefore, it is extremely narrow
with a width of order 10 keV \cite{CH,Bardeen,Colangelo}. As for
$D_{s1}(2460)$, it is below $D^*K$ threshold and its decay to $DK$
is forbidden by parity and angular momentum conservation. Hence,
the allowed strong decays are $D_s^*\pi^0$, $D_{s0}^*\pi^0$,
$D_s\pi\pi$ and $D_s\pi\pi\pi$. The isospin-violating decay
$D_{s1}(2460)\to D_s^*\pi^0$ has a rate similar to
$D_{s0}^*(2317)\to D_s\pi^0$. At a first sight, it is tempted to
argue that $D_s\pi\pi$ could dominate over $D_s^*\pi^0$ as the
former can proceed without violating isospin symmetry. However, a
detailed analysis shows this is not the case. The decay $D_{s1}\to
D_s\pi\pi$ arises from the weak transitions $D_{s1}(2460)\to D_s
\{\sigma(600), f_0(980),\cdots\}$ followed by the strong decays
$\{\sigma(600), f_0(980),\cdots\}\to \pi\pi$. Consider the
dominant contributions from the intermediate states $\sigma(600)$,
$f_0(980)$ and their mixing
 \be
 f_0 = s\bar s\,\cos\phi+n\bar n\sin\phi,  \quad
 \sigma = -s\bar s\,\sin\phi+n\bar n\cos\phi,
 \en
with $n\bar n\equiv(u\bar u+d\bar d)/\sqrt{2}$. The
$\sigma\!-\!f_0$ mixing angle can be inferred from various
processes, see \cite{ChengDSP} for a summary. In general, the
mixing angle is small so that $f_0(980)$ has a large $s\bar s$
component while $\sigma$ is primarily $n\bar n$. The $f_0$
production is favored by the weak decay of $D_{s1}$ into $D_s$,
but its contribution to $\pi\pi$ is suppressed by the large
off-shellness of $f_0(980)$, recalling that the mass difference
between $D_{s1}(2460)$ and $D_s$ is only 494 MeV. In contrast,
$\sigma(600)$ is favored by phase space consideration and yet its
contribution is suppressed by the small $\sigma-f_0$ mixing angle.
As a net result, although the strong decay into $D_s\pi\pi$ is
isospin conserving, its OZI suppression is more severe than the
isospin one for $D_s^*\pi^0$. This is confirmed by the recent
measurement of CLEO \cite{CLEO}
 \be
 {\B(D_{s1}(2460)\to D_s\pi^+\pi^-)\over \B(D_{s1}(2460)\to
 D_s^*\pi^0)}<0.08\,.
 \en
As for the electromagnetic decays of $D_{s1}(2460)$, CLEO and
Belle found
 \be \label{gammatopi0}
  {\B(D_{s1}(2460)\to D_s\gamma)\over \B(D_{s1}(2460)\to
 D_s^*\pi^0)}=\cases{
 <0.49 & CLEO \cite{CLEO}, \cr
 0.38\pm0.11\pm0.04 & Belle \cite{BelleD}, \cr
 0.63\pm0.15\pm0.15 & Belle \cite{BelleDs}.
 }
 \en
There are two different measurements of the radiative mode by
Belle: the measured value $0.38\pm0.11\pm0.04$ is determined from
$B\to DD_{s1}$ decays, while the other value comes from charm
fragmentation of $e^+e^-\to c\bar c$. These two measurements are
consistent with each other, though the central values differ
significantly. Hence, just as its $0^+$ partner $D_{s0}^*(2317)$,
$D_{s1}(2460)$ is also extremely narrow. A theoretical estimation
yields 38.2 keV for its width \cite{Bardeen}.

Eq. (\ref{h'}) leads to $\Gamma(D_{s2}^*)=12.6$ MeV, in agreement
with experiment (see Table I). Since $\Gamma(D_{s1}^{3/2})=280$
keV followed from Eqs. (\ref{Dwaverate}) and (\ref{h'}) and
$D_{s1}^{1/2}$ is very narrow as its mass is close to
$D_{s1}(2460)$ which is below $D^*K$ threshold, the decay width of
$D'_{s1}(2536)$ is thus at most of order 0.3 MeV and is consistent
with the experimental limit 2.3 MeV \cite{PDG}. In short, while
$D'_1$ and $D_2^*$ are rather narrow, $D_0^*$ and $D_1$ are quite
broad as they are allowed to have $s$-wave hadronic decays. In
sharp contrast, $D_{s0}^*$ and $D_{s1}$ are even much narrower
than $D'_{s1}$ as their allowed strong decays are isospin
violating.

Since the width of $D'_{s1}(2536)$ has not been measured, we will
appeal to the quark potential model to estimate the
$D_{s1}^{1/2}\!-\!D_{s1}^{3/2}$ mixing angle [see Eq.
(\ref{Dsmixing})]. It is known that spin-orbital and tensor force
interactions are responsible for the mass splitting of the four
$p$-wave charmed mesons. In the quark potential model the relevant
mass operator has the form \cite{Cahn}
 \be \label{massoperator}
M=\lambda \mbox{\boldmath  $\ell$}\cdot {\bf
s}_1+4\tau\mbox{\boldmath  $\ell$}\cdot{\bf s}_2 + \tau S_{12},
 \en
with ${\bf s}_1$ and ${\bf s}_2$ refer to the spin of the light
and heavy quarks, respectively, and
 \be
 \lambda &=& {1\over 2m_q^2}\left[{V'\over r}\left(1+{2m_q\over
 m_c}\right)-{S'\over r}\right], \non \\
 \tau &=& {1\over 4m_qm_c}\,{V'\over r},
 \en
where $V(r)$ is the zero-component of a vector potential, $S(r)$
is a scalar potential responsible for confinement, and $S_{12}$ is
the tensor force operator
 \be
 S_{12}=3\mbox{\boldmath  $\sigma$}_1\cdot \hat {\bf r}\,\mbox{\boldmath
   $\sigma$}_2\cdot \hat {\bf r}-\mbox{\boldmath  $\sigma$}_1\cdot\mbox{\boldmath
$\sigma$}_2.
 \en
Note that the assumption of a Coulomb-like potential for $V(r)$
has been made in deriving Eq. (\ref{massoperator}) \cite{Cahn}.
Under this hypothesis, the mass splitting is governed by the two
parameters $\lambda$ and $\tau$.

Following Cahn and Jackson \cite{Cahn}, the masses of the $J=2$
and $J=0$ states read
 \be
 M_2={\lambda\over 2}+{8\over 5}\tau+c, \qquad
 M_0=-\lambda-8\tau+c,
 \en
while the masses of the two $J=1$ states obtained by diagonalizing
the matrix in the $|J,j,m\ra=|1,3/2,m\ra$ and $|1,1/2,m\ra$ bases
are (up to a common mass $c$)
 \be \label{Mmatrix}
 \left(\matrix{
  {\lambda\over 2}-{8\over 3}\tau & -{2\sqrt{2}\over 3}\tau \cr
  -{2\sqrt{2}\over 3}\tau & -\lambda+{8\over 3}\tau \cr}\right).
 \en
It is clear that the mixing vanishes in the heavy quark limit
$m_c\to\infty$. However, $1/m_c$ corrections will allow charm
quark spin to flip and mix $D_1^{1/2}$ and $D_1^{3/2}$. The two
eigenmasses for $J=1$ are then
 \be
 M_{1\pm}=-{\lambda\over 4}+c \pm\sqrt{{\lambda^2\over 16}+{1\over
 2}(\lambda-4\tau)^2}.
 \en
From Eqs. (\ref{Dsmixing}) and (\ref{Mmatrix}) we arrive at
 \be
 \theta=\sin^{-1}\left({-R_+\over\sqrt{1+R_+^2}}\right)~~~{\rm with}~~
 R_+={{\lambda\over 2}-{8\over 3}\tau-M_{1+}\over {2\sqrt{2}\over
 3}\tau}.
 \en
The parameters $\lambda$ and $\tau$ are obtained by a global fit
to the charm spectroscopy. For $D_s^{**}$ mesons, it is found
 \be
  \tau\approx 12\,{\rm MeV}, \qquad \lambda\approx 104\,{\rm MeV}, \qquad
  \theta_s\approx 7^\circ.
  \en
As pointed out in \cite{Cahn}, a positive spin-orbit energy
$\lambda$ implies a less important scalar potential $S$. On the
contrary, the existing potential model calculation such as the one
by Di Pierro and Eichten  \cite{QM} yields $\lambda<0$ \cite{Cahn}
or a very strong confining potential $S$. This will cause a
reversed splitting, namely the $j=1/2$ states lying above the
$j=3/2$ states. However, we will not address this issue here. For
$D^{**}$ mesons, the parameters $\lambda$ and $\tau$ fall into
some large regions because of large uncertainties associated with
the measured masses of $D_0^*$ and $D_1$. Hence, the magnitude and
even the sign of the $D_1^{1/2}-D_1^{3/2}$ mixing angle $\theta$
at present cannot be fixed within this approach. Instead, we have
used heavy quark effective theory together with the measured
widths to extract $|\theta|$. As will be seen below, the sign of
$\theta$ can be inferred from a study of the $p$-wave charmed
meson production in $B$ decays.

\section{Decay constants and form factors}

\subsection{Decay constants}
The decay constants of scalar and pseudoscalar mesons are defined
by
 \be \label{decaycon}
  \la 0|A_\mu|P(q)\ra &=& if_Pq_\mu, \qquad\qquad\quad \la
0|V_\mu|S(q)\ra= f_S q_\mu.
 \en
It is known that the decay constants of non-charm light scalar
mesons are smaller than that of pseudoscalar mesons as they vanish
in the SU(3) limit. For the neutral scalars $\sigma(600)$,
$f_0(980)$ and $a_0^0(980)$, the decay constant must be zero owing
to charge conjugation invariance or conservation of vector
current:
 \be
 f_{\sigma}=f_{f_0}=f_{a_0^0}=0.
 \en
Applying the equation of motion, it is easily seen that the decay
constant of $K^{*+}_0$ ($a^+_0$) is proportional to the mass
difference between the constituent $s$ ($d$) and $u$ quarks.
Consequently, the decay constant of the charged $a_0(980)$ is very
small, while the one for $K^*_0(1430)$ is less suppressed. A
calculation based on the finite-energy sum rules \cite{Maltman}
yields
 \be
 f_{a^\pm_0}=1.1\,{\rm MeV}, \qquad f_{K^*_0}=42\,{\rm MeV}.
 \en

Contrary to the non-charm scalar resonances, the decay constant of
the scalar charmed meson is not expected to be suppressed because
of charm and light quark mass imbalance. Applying the equation of
motion again leads to
 \be
 m_{K^*_0}^2f_{K^*_0} = i(m_s-m_u)\la K^*_0|\bar su|0\ra, \qquad\qquad
 m_{D^*_0}^2f_{D^*_0} = i(m_c-m_u)\la D^*_0|\bar cu|0\ra.
 \en
For a crude estimate, we assume $\la D^*_0|\bar c u|0\ra\approx
\la K^*_0|\bar s u|0\ra$ and obtain
 \be \label{fD0}
 f_{D^*_0}\approx 160\,{\rm MeV}.
 \en
This is comparable to $f_{D}\approx 200$ MeV, the decay constant
of the pseudoscalar $D$ meson.

The decay constants of the axial-vector charmed mesons are defined
by
 \be
 \la 0|A_\mu|D_1^{1/2}(q,\vp)\ra=\,f_{D_1^{1/2}}m_{D_1^{1/2}}\vp_\mu, \qquad
  \la 0|A_\mu|D_1^{3/2}(q,\vp)\ra=\,f_{D_1^{3/2}}m_{D_1^{3/2}}\vp_\mu.
 \en
It has been shown that in the heavy quark limit
\cite{Colangelo92,Yaouanc,Veseli}
 \be \label{fHQS}
 f_{D_1^{1/2}}=f_{D_0^*}, \qquad\quad f_{D_1^{3/2}}=0.
 \en
Since the decay constant of $D_2^*$ vanishes irrespective of heavy
quark symmetry (see below), the charmed mesons within the
multiplet $(0^+,1^+)$ or $(1'^+,2^+)$ thus have the same decay
constant. This is opposite to the case of light $p$-wave mesons
where the decay constant of $^1P_1$ meson vanishes in the SU(3)
limit \cite{Suzuki} based on the argument that for non-charm axial
vector mesons, the $^3P_1$ and $^1P_1$ states transfer under
charge conjunction as
 \be
 M_a^b(^3P_1) \to M_b^a(^3P_1), \qquad M_a^b(^1P_1) \to
 -M_b^a(^1P_1),~~~(a=1,2,3),
 \en
where the axial-vector mesons are represented by a $3\times 3$
matrix. Since the weak axial-vector current transfers as
$(A_\mu)_a^b\to (A_\mu)_b^a$ under charge conjugation, it is clear
that the decay constant of the $^1P_1$ meson vanishes in the SU(3)
limit \cite{Suzuki}.

The polarization tensor $\vp_{\mu\nu}$ of a tensor meson satisfies
the relations
 \be
 \vp_{\mu\nu}=\vp_{\nu\mu}, \qquad \vp^{\mu}_{~\mu}=0, \qquad p_\mu
 \vp^{\mu\nu}=p_\nu\vp^{\mu\nu}=0.
 \en
Therefore,
 \be
 \la 0|(V-A)_\mu|D^*_2(\vp,p)\ra=a\vp_{\mu\nu}p^\nu+b\vp^\nu_{~\nu}
 p_\mu=0.
 \en
The above relation in general follows from Lorentz covariance and
parity consideration. Hence the decay constant of the tensor meson
vanishes; that is, the tensor meson $D_2^*$ cannot be produced
from the $V-A$ current.

Beyond the heavy quark limit, the relations (\ref{fHQS}) receive
large $1/m_c$ corrections which have been estimated in
\cite{Veseli} using the relativistic quark model. In the present
paper we shall use $f_\rho=216$ MeV and (in units of MeV)
 \be \label{decayconst}
&& f_{D}=200, \qquad~~ f_{D_s}=230, \qquad~~ f_{D_s^*}=230, \non \\
&& f_{D_0}=160, \qquad~ f_{D_1^{1/2}}=120, \qquad f_{D_1^{3/2}}=40, \non \\
&& f_{D_{s0}}=60, \qquad f_{D_{s1}^{1/2}}=145, \qquad
f_{D_{s1}^{3/2}}=50.
 \en
The decay constants for $D_s^{**}$ are obtained essentially by
fitting to experiment to be discussed in Sec. IV.B. Note that the
measurements of $B\to D_s^{(*)}D^{(*)}$ \cite{PDG,BaBarDs}
indicate that the decay constants of $D_s^*$ and $D_s$ are
similar.

\subsection{Form factors}

Form factors for $B\to M$ transitions with $M$ being a parity-odd
meson are given by \cite{BSW}
  \be
 \la P(p)|V_\mu|B(p_B)\ra &=& \left((p_B+p)_\mu-{m_B^2-m_{P}^2\over q^2}\,q_ \mu\right)
F_1^{BP}(q^2)+{m_B^2-m_{P}^2\over q^2}q_\mu\,F_0^{BP}(q^2), \non \\
\la V(p,\vp)|V_\mu|B(p_B)\ra &=& {2\over
m_P+m_V}\,\epsilon_{\mu\nu\alpha \beta}\vp^{*\nu}p_B^\alpha
p^\beta  V(q^2),   \non \\
 \la V(p,\vp)|A_\mu|B(p_B)\ra &=& i\Big\{
(m_P+m_V)\vp^*_\mu A_1(q^2)-{\vp^*\cdot p_B\over
m_P+m_V}\,(p_B+p)_\mu A_2(q^2)    \non \\
&& -2m_V\,{\vp^*\cdot p_B\over
q^2}\,q_\mu\big[A_3(q^2)-A_0(q^2)\big]\Big\},
 \en
where $q=p_B-p$, $F_1(0)=F_0(0)$, $A_3(0)=A_0(0)$, and
 \be
A_3(q^2)=\,{m_P+m_V\over 2m_V}\,A_1(q^2)-{m_P-m_V\over
2m_V}\,A_2(q^2).
 \en
For $B\to P$ and $B\to V$ form factors, we will use the
Melikhov-Stech (MS) model \cite{MS} based on the constituent quark
picture. Other form factor models give similar results.

The general expressions for $B\to D^{**}$ transitions ($D^{**}$
being a $p$-wave charmed meson) are given by \cite{ISGW}
 \be \label{ISGWform}
 \la D^*_0(p)|A_\mu|B(p_B)\ra &=& i\Big[ u_+(q^2)(p_B+p)_\mu+u_-(q^2)(p_B-p)_\mu
 \Big], \non \\
 \la D_1^{1/ 2}(p,\vp)|V_\mu|B(p_B)\ra &=& i\left[\ell_{1/2}(q^2)\vp_\mu^*+c_+^{1/2}(q^2)(\vp^*\cdot
 p_B)(p_B+p)_\mu+c_-^{1/2}(q^2)(\vp^*\cdot p_B)(p_B-p)_\mu\right], \non \\
 \la D_1^{1/2}(p,\vp)|A_\mu|B(p_B)\ra &=&
 -q_{1/2}(q^2)\epsilon_{\mu\nu\rho\sigma}\vp^{*\nu}(p_B+p)^\rho(p_B-p)^\sigma,
 \non \\
 \la D_1^{3/ 2}(p,\vp)|V_\mu|B(p_B)\ra &=& i\left[\ell_{3/2}(q^2)\vp_\mu^*+c_+^{3/2}(q^2)(\vp^*\cdot
 p_B)(p_B+p)_\mu+c_-^{3/2}(q^2)(\vp^*\cdot p_B)(p_B-p)_\mu\right], \non \\
 \la D_1^{3/2}(p,\vp)|A_\mu|B(p_B)\ra &=&
 -q_{3/2}(q^2)\epsilon_{\mu\nu\rho\sigma}\vp^{*\nu}(p_B+p)^\rho(p_B-p)^\sigma,
 \non \\
 \la D^*_2(p_,\vp)|V_\mu|B(p_B)\ra &=&
 h(q^2)\epsilon_{\mu\nu\rho\sigma}\vp^{*\nu\alpha}(p_B)_\alpha(p_B+p)^\rho
 (p_B-p)^\sigma,  \non \\
 \la D^*_2(p,\vp)|A_\mu|B(p_B)\ra &=& -i\Big[k(q^2)\vp^*_{\mu\nu}p_B^\nu+
 b_+(q^2)\vp^*_{\alpha\beta}p_B^\alpha p_B^\beta(p_B+p)_\mu
 \non \\
 &+& b_-(q^2)\vp^*_{\alpha\beta}p_B^\alpha
 p_B^\beta(p_B-p)_\mu\Big].
 \en
In order to know the sign of various form factors appearing in Eq.
(\ref{ISGWform}), it is instructive to check the heavy quark limit
behavior of $B\to D^{**}$ transitions which has the form \cite{IW}
 \be \label{formHQS}
 \la  D_0^*(v')|A_\mu|B(v)\ra &=& \sqrt{m_Bm_{D_0}}\,2\tau_{1/2}(\omega)i(v'-v)_\mu,
 \non \\
 \la D_1^{1/2}(v',\vp)|V_\mu|B(v)\ra &=& \sqrt{m_Bm_{D_1^{1/2}}}\,2\tau_{1/2}(\omega)
 i\Big[ (\omega-1)\vp^*_\mu-(\vp^*\cdot v)v'_\mu\Big], \non \\
 \la D_1^{1/2}(v',\vp)|A_\mu|B(v)\ra &=& \sqrt{m_Bm_{D_1^{1/2}}}\,2\tau_{1/2}(\omega)
 (-)\epsilon_{\mu\nu\alpha\beta} \vp^{*\nu}{v'}^\alpha v^\beta, \non \\
 \la D_1^{3/2}(v',\vp)|V_\mu|B(v)\ra &=& \sqrt{ {1\over 2}m_Bm_{D_1^{3/2}}}\,\tau_{3/2}(\omega)
 i\Big\{ (1-\omega^2)\vp^*_\mu-(\vp^*\cdot v)[3v_\mu-(\omega-2)v'_\mu]\Big\}, \non \\
 \la D_1^{3/2}(v',\vp)|A_\mu|B(v)\ra &=& \sqrt{{1\over 2}m_Bm_{D_1^{3/2}}}\,\tau_{3/2}(\omega)
 (\omega+1)\epsilon_{\mu\nu\alpha\beta} \vp^{*\nu}{v'}^\alpha v^\beta,  \non \\
 \la D^*_2(v',\vp)|V_\mu|B(v)\ra &=& \sqrt{3m_Bm_{D_2}}\,\tau_{3/2}(\omega)
 \epsilon_{\mu\nu\alpha\beta} \vp^{*\nu\gamma}v_\gamma {v'}^\alpha v^\beta, \non \\
 \la D^*_2(v',\vp)|A_\mu|B(v)\ra &=& \sqrt{3m_Bm_{D_2}}\,\tau_{3/2}(\omega)
 (-i)\Big\{ (\omega+1)\vp^*_{\mu\nu}v^\nu-\vp^*_{\alpha\beta}v^\alpha v^\beta
 v'_\mu\Big\},
 \en
where $\omega\equiv v\cdot v'$ and there are two independent
functions $\tau_{1/2}(\omega)$ and $\tau_{3/2}(\omega)$ first
introduced in \cite{IW}. It is easily seen that the matrix
elements of weak currents vanish at the zero recoil point
$\omega=1$ owing to the orthogonality of the wave functions of $B$
and $D^{**}$. The universal functions $\tau_{1/2}(\omega)$ and
$\tau_{3/2}(\omega)$ are conventionally parametrized as
 \be
 \tau_i(\omega)=\tau_i(1)[1-\rho^2_i(\omega-1)]
 \en
for $i=1/2$ and $3/2$. The slope parameter $\rho^2$ can be related
to $\tau_{1/2}(1)$ and $\tau_{3/2}(1)$ via the Bjorken sum rule.
Comparing Eq. (\ref{ISGWform}) with Eq. (\ref{formHQS}) we see
that heavy quark symmetry requires that the form factors
$u_+,\ell_{1/2},q_{1/2},c_-^{1/2},h,k$ and $b_-$ be positive,
while $u_-,\ell_{3/2}, q_{3/2},c_+^{1/2}, c_+^{3/2},c_-^{3/2}$ and
$b_+$ be negative. Heavy quark symmetry also demands the relations
$c_+^{1/2}+c_-^{1/2}=0$ and $b_++b_-=0$. It is easily seen that
these heavy quark symmetry requirements are satisfied in realistic
model calculations shown below.

In the present paper, we shall use the improved version, the
so-called ISGW2 model \cite{ISGW2}, of the non-relativistic quark
model by Isgur-Scora-Grinstein-Wise (ISGW) \cite{ISGW} to compute
the $B\to D^{**}$ transition form factors.\footnote{Note that in
the original version of the ISGW model \cite{ISGW}, the form
factors for $B$ to axial-vector charmed meson transition are
evaluated for $D_1(^1P_1)$ and $D_1(^3P_1)$. As a result, one has
to apply Eq. (\ref{PjJ}) to obtain $B\to D_1^{1/2}$ and $B\to
D_1^{3/2}$ form factors.} In general, the form factors evaluated
in the original version of the ISGW model are reliable only at
$q^2=q^2_m$, the maximum momentum transfer. The reason is that the
form-factor $q^2$ dependence in the ISGW model is proportional to
exp[$-(q^2_m-q^2)$] and hence the form factor decreases
exponentially as a function of $(q^2_m-q^2)$. This has been
improved in the ISGW2 model in which the form factor has a more
realistic behavior at large $(q^2_m-q^2)$ which is expressed in
terms of a certain polynomial term. In addition to the form-factor
momentum dependence, the ISGW2 model incorporates a number of
improvements, such as the constraints imposed by heavy quark
symmetry, hyperfine distortions of wave functions, etc.,$\cdots$
\cite{ISGW2}. The results of the ISGW2 model predictions for
various form factors are shown in Tables
\ref{tab:BtoD0}-\ref{tab:BtoD2}. Evidently, the signs of various
calculated form factors are consistent with what are expected from
heavy quark symmetry.

\begin{table}[h]
\caption{The form factors at various $q^2$ for $B\to D_0^*$ and
$B_s\to D_{s0}^*$ transitions calculated in the ISGW2 model.
 } \label{tab:BtoD0}
\begin{center}
\begin{tabular}{l c c c c c c }
Transition & $u_+(m_\pi^2)$ & $u_-(m_\pi^2)$ & $u_+(m_\rho^2)$ &
$u_-(m_\rho^2)$ & $u_+(m_{D_s}^2)$ & $u_-(m_{D_s}^2)$ \\
\hline
 $B\to D_0^*$ & 0.175 & $-0.462$ & 0.178 & $-0.471$ &  0.198 & $-0.524$ \\
 $B_s\to D_{s0}^*$ & 0.196 & $-0.515$ & 0.200 & $-0.527$ & 0.230 & $-0.605$  \\
\end{tabular}
\end{center}
\end{table}

\begin{table}[h]
\caption{The form factors at $q^2=m_\pi^2$ for $B\to D_1^{1/2}$
and $B\to D_1^{3/2}$ transitions calculated in the ISGW2 model,
where $\ell_{1/2}$ and $\ell_{3/2}$ are in units of GeV and all
others are in units of ${\rm GeV}^{-1}$.
 } \label{tab:BtoD1}
\begin{center}
\begin{tabular}{l c c c c c c c c }
Transition & $q_{1/2}$ & $\ell_{1/2}$ & $c_+^{1/2}$ & $c_-^{1/2}$
& $q_{3/2}$ & $\ell_{3/2}$ & $c_+^{3/2}$ & $c_-^{3/2}$ \\
\hline
 $B\to D_1^{1/2}$ & $0.057$ & $0.54$ & $-0.064$ & $0.068$ & & &  \\
 $B\to D_1^{3/2}$ & & & & & $-0.057$ & $-1.15$ & $-0.043$ & $-0.018$ \\
 $B_s\to D_{s1}^{1/2}$ & $0.063$ & $0.66$ & $-0.072$ & $0.078$ & & &  \\
 $B_s\to D_{s1}^{3/2}$ & & & & & $-0.063$ & $-1.31$ & $-0.048$ & $-0.023$ \\
\end{tabular}
\end{center}
\end{table}

\begin{table}[h]
\caption{The $B\to D_2^*$ and $B_s\to D_{s2}^*$ form factors at
$q^2=m_\pi^2$ calculated in the ISGW2 model, where $k$ is
dimensionless and $h$, $b_+$, $b_-$ are in units of ${\rm
GeV}^{-2}$.
 } \label{tab:BtoD2}
\begin{center}
\begin{tabular}{l c c c c }
Transition & $h$ & $k$ & $b_+$ & $b_-$ \\
\hline
 $B\to D^*_2$ & 0.011 & 0.60 & $-0.010$ & 0.010 \\
 $B_s\to D^*_{s2}$ & 0.013 & 0.70 & $-0.011$ & 0.012 \\
\end{tabular}
\end{center}
\end{table}

\begin{table}[h]
\caption{The form factors $F_0$ and $F_1$ at various $q^2$ for
$B\to D_0^*$ and $B_s\to D_{s0}^*$ transitions calculated in the
ISGW2 model.
 } \label{tab:BtoD0a}
\begin{center}
\begin{tabular}{l c c c c c c }
Transition & $F_1(m_\pi^2)$ & $F_0(m_\pi^2)$ & $F_1(m_\rho^2)$ &
$F_0(m_\rho^2)$
& $F_1(m_{D_s}^2)$ & $F_0(m_{D_s}^2)$ \\
\hline
 $B\to D_0^*$ & 0.175 & 0.175 & 0.178 & 0.166 &  0.198 & 0.108 \\
 $B_s\to D_{s0}^*$ & 0.196 & 0.196 & 0.200 & 0.187 & 0.230 & 0.130  \\
\end{tabular}
\end{center}
\end{table}

\begin{table}[h]
\caption{The dimensionless form factors $A$ and $V_{0,1,2}$ at
$q^2=m_\pi^2$ for $B\to D_1^{1/2}$ and $B\to D_1^{3/2}$
transitions calculated in the ISGW2 model.
 } \label{tab:BtoD1a}
\begin{center}
\begin{tabular}{l r r r r }
Transition & $A$ & $V_0$ & $V_1$ & $V_2$ \\
\hline
 $B\to D_1^{1/2}$ & $-0.43$ & $-0.18$ & $0.070$ & $0.49$  \\
 $B\to D_1^{3/2}$ & $0.44$ & $-0.43$ & $-0.15$ & $0.33$ \\
 $B_s\to D_{s1}^{1/2}$ & $-0.49$ & $-0.20$ & $0.085$ & $0.57$ \\
 $B_s\to D_{s1}^{3/2}$ & $0.50$ & $-0.47$ & $-0.17$ & $0.38$ \\
\end{tabular}
\end{center}
\end{table}

In realistic calculations of decay amplitudes it is convenient to
employ the dimensionless form factors defined by \cite{BSW}
 \be
 \la D^*_0(p_{D_0})|A_\mu|B(p_B)\ra &=&
i\left[\left( (p_B+p_{D_0})_\mu-{m_B^2-m_{D_0}^2\over q^2}\,q_
\mu\right) F_1^{BD_0}(q^2)+{m_B^2-m_{D_0}^2\over
q^2}q_\mu\,F_0^{BD_0}(q^2)\right], \non \\
 \la D_1(p_{D_1},\vp)|V_\mu|B(p_B)\ra &=&
i\Bigg\{(m_B+m_{D_1}) \vp^*_\mu V_1^{BD_1}(q^2)  - {\vp^*\cdot
p_B\over m_B+m_{D_1}}(p_B+p_{D_1})_\mu V_2^{BD_1}(q^2) \non \\
&-& 2m_{D_1} {\vp^*\cdot p_B\over
q^2}(p_B-p_{D_1})_\mu\left[V_3^{BD_1}(q^2)-V_0^{BD_1}(q^2)\right]\Bigg\},
\non \\
  \la D_1(p_{D_1},\vp)|A_\mu|B(p_B)\ra &=& {2\over
  m_B+m_{D_1}}\epsilon_{\mu\nu\rho\sigma}\vp^{*\nu}p_B^\rho p_{D_1}^\sigma
  A^{BD_1}(q^2),
 \en
with
 \be V_3^{BD_1}(q^2)=\,{m_B+m_{D_1}\over 2m_{D_1}}\,V_1^{BD_1}(q^2)-{m_B-m_{D_1}\over
2m_{D_1}}\,V_2^{BD_1}(q^2),
 \en
and $V_3^{BD_1}(0)=V_0^{BD_1}(0)$. The form factors relevant for
$B\to D_0^*P$ decays are $F_0^{BD_0}$ and $F_0^{BP}$. Note that
only the form factor $V_0^{BP}$ or $F_1^{BD_1}$ will contribute to
the factorizable amplitude of $B\to D_1P$ as one can check the
matrix elements $q^\mu \la D_1(p_{D_1},\vp)|V_\mu|B(p_B)\ra$ and
$\vp^\mu\la P|V_\mu|B\ra$. The ISGW2 model predictions for the
form factors $F_{0,1}$, $V_{0,1,2}$ and $A$ are summarized in
Tables \ref{tab:BtoD0a}-\ref{tab:BtoD1a}. It is evident that the
form factor $F_{0,1}(0)\approx 0.18$ for $B\to D_0^*$ transition
is much smaller than the typical value of $0.65\sim 0.70$ for the
$B\to D$ transition form factor at $q^2=0$.

\section{Analysis of $\ov B\to D^{**}M,~\ov D_{\lowercase{s}}^{**}M$ Decays}

\subsection{Factorization}

In the present work we focus on the Cabibbo-allowed decays  $\ov
B\to D^{**}\pi(\rho),~D^{**}\bar D_s^{(*)}$, $\bar
D^{**}_sD^{(*)}$ and $\ov B_s\to D_s^{**}\pi(\rho)$, where
$D^{**}$ denotes generically a $p$-wave charmed meson. We will
study these decays within the framework of generalized
factorization in which the hadronic decay amplitude is expressed
in terms of factorizable contributions multiplied by the {\it
universal} (i.e. process independent) effective parameters $a_i$
that are renormalization scale and scheme independent. Since the
aforementioned $B$ decays either proceed through only via tree
diagrams or are tree dominated, we will thus neglect the small
penguin contributions and write the weak Hamiltonian in the form
 \be
 H_{\rm eff} &=& {G_F\over\sqrt{2}}\Bigg\{ V_{cb}V_{cs}^*\Big[ a_1(\bar cb)
 (\bar sc)+a_2(\bar sb)(\bar cc)\Big]
 + V_{cb}V_{ud}^*\Big[ a_1(\bar cb)(\bar du)+a_2(\bar db)(\bar
 cu)\Big]+h.c.
 \en
with $(\bar q_1q_2)\equiv \bar q_1\gamma_\mu(1-\gamma_5)q_2$. For
hadronic $B$ decays, we shall use $a_1=1.15$ and $a_2=0.26$\,.

Under the factorization hypothesis, the decays $B^-\to D^{**0}\bar
D_s^{-}$, $\ov B^0\to D^{**+}\bar D_s^{-}$ and $\ov B_s^0\to
D_s^{**+}\pi^-$ receive contributions only from the external
$W$-emission diagram. As stated before, the penguin contributions
to the first two decay modes are negligible.

Apart from a common factor of $G_FV_{cb}V_{ud}^*/\sqrt{2}$, the
factorizable amplitudes for $B^-\to D^{**0}\pi^-$ read
 \be \label{BtoDpi}
 A(B^-\to D_0^*(2308)^0\pi^-) &=& -a_1
 f_\pi(m_B^2-m_{D_0}^2)F_0^{BD_0}(m_\pi^2)
 -a_2 f_{D_0}(m_B^2-m_\pi^2)F_0^{B\pi}(m_{D_0}^2), \non \\
 A(B^-\to D_1(2427)^0\pi^-) &=&
 -2(\vp^*\cdot p_B)\Bigg\{a_1f_\pi\Big[V_0^{BD_1^{3/2}}(m_\pi^2)m_{D_1^{3/2}}\sin\theta+
 V_0^{BD_1^{1/2}}(m_\pi^2)m_{D_1^{1/2}}\cos\theta\Big] \non \\
 &+& a_2\Big[F_1^{B\pi}(m_{D_1^{3/2}}^2)m_{D_1^{3/2}}f_{D_1^{3/2}}\sin\theta+
 F_1^{B\pi}(m_{D_1^{1/2}}^2)m_{D_1^{1/2}}f_{D_1^{1/2}}\cos\theta\Big]\Bigg\}, \non \\
  A(B^-\to D'_1(2420)^0\pi^-) &=&
 -2(\vp^*\cdot p_B)\Bigg\{a_1f_\pi\Big[V_0^{BD_1^{3/2}}(m_\pi^2)m_{D_1^{3/2}}\cos\theta-
 V_0^{BD_1^{1/2}}(m_\pi^2)m_{D_1^{1/2}}\sin\theta\Big] \non \\
 &+& a_2\Big[F_1^{B\pi}(m_{D_1^{3/2}}^2)m_{D_1^{3/2}}f_{D_1^{3/2}}\cos\theta-
 F_1^{B\pi}(m_{D_1^{1/2}}^2)m_{D_1^{1/2}}f_{D_1^{1/2}}\sin\theta\Big]\Bigg\}, \non \\
 A(B^-\to D^*_2(2460)^0\pi^-) &=& ia_1f_\pi\,\vp^*_{\mu\nu}p_B^\mu p_B^\nu\,\left[
 k(m_\pi^2)+b_+(m_\pi^2)(m_B^2-m_{D_2}^2)+b_-(m_\pi^2)m_\pi^2\right].
 \en
Note that except $B^-\to D_2^{*0}\pi^-$ all other modes receive
contributions from color-suppressed internal $W$-emission.  The
decay rates are given by
 \be
 \Gamma(B\to D_0^*\pi) &=& {p_c\over 8\pi m_B^2}|A(B\to
 D_0^*\pi)|^2, \non \\
 \Gamma(B\to D_1\pi) &=& {p^3_c\over 8\pi m_{D_1}^2}|A(B\to
 D_1\pi)/(\vp^*\cdot p_B)|^2, \non \\
 \Gamma(B\to D_2^*\pi) &=& {p_c^5\over 12\pi m_{D_2}^2}\left({m_B\over
 m_{D_2}}\right)^2|M(B\to D_2^*\pi)|^2,
 \en
where $A(B\to D^*_2\pi)= \vp^*_{\mu\nu}p_B^\mu p_B^\nu\,M(B\to
D^*_2\pi)$ and $p_c$ is the c.m. momentum of the pion. The
$p_c^{2L+1}$ dependence in the decay rate indicates that only
$s$-, $p$- and $d$-waves are allowed in $D_0^*\pi$, $D_1\pi$ and
$D_2^*\pi$ systems, respectively. The factorizable  decay
amplitudes for $B^-\to D_0^{*0}\rho^-$ and $B^-\to D_2^{*0}\rho^-$
are (up to a common factor of $G_FV_{cb}V_{ud}^*/\sqrt{2}$)
 \be
 A(B^-\to D_0^*(2308)^0\rho^-) &=& 2(\vp^*\cdot p_B)\left[a_1f_\rho m_\rho
 F_1^{BD_0}(m_\rho^2)+a_2 f_{D_0}m_{D_0}A_0^{B\rho}(m_{D_0}^2)\right],
 \non \\
 A(B^-\to D_2^*(2460)^0\rho^-) &=& a_1f_\rho m_\rho^2
 \vp^{*\alpha\beta}\vp_\mu^*(p_B-p_{D_2})_\lambda\Big[ih(m_\rho^2)\epsilon^{\mu\nu\lambda\sigma}
 g_{\alpha\nu}(p_\rho)_\beta(p_\rho)_\sigma \non \\
 &+& k(m_\rho^2)\delta^\mu_\alpha\delta^\lambda_\beta+b_+(m_\rho^2)(p_\rho)_\alpha
 (p_\rho)_\beta g^{\mu\lambda}\Big].
 \en

The expression for $\ov B\to D_1\rho$ is more complicated. In the
absence of the $D_1^{1/2}-D_1^{3/2}$ mixing, one has
 \be
 A(B^-\to D'_1(2420)^0\rho^-) &=& -ia_1f_\rho m_\rho \Big[(\vp^*_\rho\cdot\vp^*_{D_1})
(m_{B}+m_{D_1})V_1^{ BD_1}(m_\rho^2)  \non \\
&-& (\vp^*_\rho\cdot p_{_{B}})(\vp^*_{D_1} \cdot p_{_{B}}){2V_2^{
BD_1}(m_{\rho}^2)\over m_{B}+m_{D_1} } +
i\epsilon_{\mu\nu\alpha\beta}\vp^{*\mu}_{D_1}\vp^{*\nu}_\rho
p^\alpha_{_{B}} p^\beta_\rho\,{2A^{BD_1}(m_{\rho}^2)\over
m_{B}+m_{D_1} }\Bigg] \non \\
&-& ia_2f_{D_1}m_{D_1} \Big[(\vp^*_\rho\cdot\vp^*_{D_1})
(m_{B}+m_\rho)A_1^{ B\rho}(m_{D_1}^2)  \non \\
&-& (\vp^*_\rho\cdot p_{_{B}})(\vp^*_{D_1} \cdot p_{_{B}}){2A_2^{
B\rho}(m_{D_1}^2)\over m_{B}+m_{\rho} } +
i\epsilon_{\mu\nu\alpha\beta}\vp^{*\mu}_{\rho}\vp^{*\nu}_{D_1}
p^\alpha_{_{B}} p^\beta_{D_1}\,{2V^{B\rho}(m_{D_1}^2)\over
m_{B}+m_{\rho} }\Bigg].
 \en
In the presence of the $D_1^{1/2}-D_1^{3/2}$ mixing, it is more
convenient to express the decay amplitude as
 \be
 A[B^-\to D_1^0(\vp_{D_1},p_{D_1})\rho^-(\vp_\rho,p_\rho)] \propto
\vp_{D_1}^{*\mu}\vp_\rho^{*\nu}[S_1 g_{\mu\nu}+S_2(p_B)_\mu
(p_B)_\nu+iS_3\epsilon_{\mu\nu\alpha\beta}p_{D_1}^\alpha
p_\rho^\beta], \label{amp}
 \en
where $\epsilon^{0123}=+1$ in our convention, the coefficient
$S_3$ corresponds to the $p$-wave amplitude, and $S_1$, $S_2$ to
the mixture of $s$- and $d$-wave amplitudes:
 \be \label{S123}
 S_1 &=& a_1f_\rho m_\rho \Big[(m_B + m_{D_1^{3/2}})V_1^{BD_1^{3/2}}(m_\rho^2)\cos\theta
 -(m_B+m_{D_1^{1/2}})V_1^{BD_1^{1/2}}(m_\rho^2)\sin\theta\Big]  \non \\
 &+& a_2(m_B+m_\rho)\Big[m_{D_1^{3/2}}f_{D_1^{3/2}}
 A_1^{B\rho}(m_{D_1^{3/2}}^2)\cos\theta-m_{D_1^{1/2}}f_{D_1^{1/2}}
 A_1^{B\rho}(m_{D_1^{1/2}}^2)\sin\theta\Big], \non \\
 S_2 &=& a_1f_\rho m_\rho \Big[{1\over m_B + m_{D_1^{3/2}}}V_2^{BD_1^{3/2}}(m_\rho^2)\cos\theta
 -{1\over m_B + m_{D_1^{1/2}}}V_2^{BD_1^{1/2}}(m_\rho^2)\sin\theta \Big]  \non \\
 &+& a_2{1\over m_B+m_\rho}\Big[m_{D_1^{3/2}}f_{D_1^{3/2}}
 A_2^{B\rho}(m_{D_1^{3/2}}^2)\cos\theta-m_{D_1^{1/2}}f_{D_1^{1/2}}
 A_2^{B\rho}(m_{D_1^{1/2}}^2)\sin\theta\Big], \non \\
 S_3 &=& a_1f_\rho m_\rho \Big[{1\over m_B + m_{D_1^{3/2}}}A^{BD_1^{3/2}}(m_\rho^2)\cos\theta
 -{1\over m_B + m_{D_1^{1/2}}}A^{BD_1^{1/2}}(m_\rho^2)\sin\theta \Big]  \non \\
 &+& a_2{1\over m_B+m_\rho}\Big[m_{D_1^{3/2}}f_{D_1^{3/2}}
 V^{B\rho}(m_{D_1^{3/2}}^2)\cos\theta-m_{D_1^{1/2}}f_{D_1^{1/2}}
 V^{B\rho}(m_{D_1^{1/2}}^2)\sin\theta\Big].
 \en
Then the helicity amplitudes $H_0$, $H_+$ and $H_-$ can be
constructed as
 \be
 H_0 &=& {1\over 2m_{D_1}m_\rho}\left[
 (m_B^2-m_{D_1}^2-m_\rho^2)S_1+2m_B^2p_c^2S_2\right], \non \\
 H_\pm &=& S_1\pm m_Bp_cS_3.
 \en
For $B^-\to D_1(2427)^0\rho^-$, the amplitudes $S_{1,2,3}$ are the
same as Eq. (\ref{S123}) except for the replacement of
$\cos\theta\to \sin\theta$ and $\sin\theta\to -\cos\theta$. The
decay rates read (up to the common factor of
$G_F^2|V_{cb}V^*_{ud}|^2/2$)
 \be
 \Gamma(B\to D_0^*\rho) &=& {p^3_c\over 8\pi m_{D_0}^2}|A(B\to
 D_0^*\rho)/(\vp^*\cdot p_B)|^2, \non \\
 \Gamma(B\to D_1\rho) &=& {p_c\over 8\pi
 m_B^2}(|H_0|^2+|H_+|^2+|H_-|^2), \non \\
 \Gamma(B\to D_2^*\rho) &=& {f_\rho^2\over 24\pi
 m_{D_2}^4}(ap_c^7+bp_c^5+cp_c^3),
 \en
with
 \be
 &&  a=8m_B^4b_+^2, \qquad\qquad c=5m_{D_2}^2m_\rho^2k^2, \non \\
 &&
 b=2m_B^2[6m_\rho^2m_{D_2}^2h^2+2(m_B^2-m_{D_2}^2-m_\rho^2)kb_++k^2].
 \en

\begin{table}[p]
\caption{The predicted branching ratios for $B^-\to
D^{**0}(\pi^-,\rho^-,\bar D_s)$ and $B^-\to \bar D_s^{**}D^{(*)}$
decays, where $D^{**}$ denotes a generic $p$-wave charmed meson.
Experimental results are taken from PDG [15] and Belle [14]. The
numbers in parentheses for $\ov D_{s0}^*D$ and $\ov D_{s1}D$
productions are fits to the central values of the experimental
data. The axial-vector meson mixing angles are taken to be
$\theta=17^\circ$ for $D_1-D'_1$ systems and $\theta_s=7^\circ$
for $D_{s1}-D'_{s1}$ systems. The parameters $a_{1,2}$ are taken
to be $a_1=1.15$ and $a_2=0.26$.
 } \label{tab:chargedB}
\begin{center}
\begin{tabular}{l c c c c l  }
Decay & This Work & KV \cite{Katoch} &  CM \cite{Castro} & KLO \cite{Kim}& Expt \\
\hline
 $B^-\to D^*_0(2308)^0\pi^-$ & $7.7\times 10^{-4}$ & $4.2\times 10^{-4}$ & &
 & $(9.2\pm2.9)\times 10^{-4}$\cite{BelleD} \\
 $B^-\to D_1(2427)^0\pi^-$ & $3.6\times 10^{-4}$ & $2.4\times 10^{-4}$ & & &
 $(7.5\pm1.7)\times 10^{-4}$\cite{BelleD} \\
 $B^-\to D'_1(2420)^0\pi^-$ & $1.1\times 10^{-3}$ & $2.1\times 10^{-3}$ & &
 & $(1.0\pm0.2)\times 10^{-3}$\cite{BelleD} \\
 & & & & & $(1.5\pm0.6)\times 10^{-3}$\cite{PDG} \\
 $B^-\to D^*_2(2460)^0\pi^-$ & $6.7\times 10^{-4}$ & $7.2\times 10^{-5}$
 & $4.1\times 10^{-4}$ & $3.5\times 10^{-4}$ & $(7.8\pm1.4)\times 10^{-4}$\cite{BelleD} \\
 \hline
 $B^-\to D^*_0(2308)^0\rho^-$ & $1.3\times 10^{-3}$ &  \\
 $B^-\to D_1(2427)^0\rho^-$ & $1.1\times 10^{-3}$ & \\
 $B^-\to D'_1(2420)^0\rho^-$ & $2.8\times 10^{-3}$ & & & & $<1.4\times 10^{-3}$\cite{PDG} \\
 $B^-\to D^*_2(2460)^0\rho^-$ & $1.8\times 10^{-3}$ & & $1.1\times 10^{-3}$
 & $9.8\times 10^{-4}$ & $<4.7\times 10^{-3}$\cite{PDG} \\
 \hline
 $B^-\to D^*_0(2308)^0\bar D_s^-$ & $8.0\times 10^{-4}$ & $2.7\times 10^{-3}$ & &\\
 $B^-\to D_1(2427)^0\bar D_s^-$ & $9.6\times 10^{-4}$ & $1.4\times 10^{-3}$ & & \\
 $B^-\to D'_1(2420)^0\bar D_s^-$ & $1.3\times 10^{-3}$ & $5.0\times 10^{-3}$ & \\
 $B^-\to D^*_2(2460)^0\bar D_s^-$ & $4.2\times 10^{-4}$ & $1.0\times 10^{-4}$
 & $2.7\times 10^{-4}$ & $4.9\times 10^{-4}$ &  \\
 \hline
 $B^-\to D^*_0(2308)^0\bar D_s^{*-}$ & $3.5\times 10^{-4}$ & & &\\
 $B^-\to D_1(2427)^0\bar D_s^{*-}$ & $6.0\times 10^{-4}$ & & & \\
 $B^-\to D'_1(2420)^0\bar D_s^{*-}$ & $1.6\times 10^{-3}$ & & \\
 $B^-\to D^*_2(2460)^0\bar D_s^{*-}$ & $1.1\times 10^{-3}$
 & $1.0\times 10^{-4}$ & $1.1\times 10^{-3}$ & $1.2\times 10^{-3}$ \\
 \hline
 $B^-\to \ov D^*_{s0}(2317)^-D^0$ & ($9.3\times 10^{-4}$) & 0 & & & see text \\
 $B^-\to \ov D_{s1}(2460)^-D^0$ &  ($3.1\times 10^{-3}$) & $3.5\times 10^{-3}$
 & & & see text \\
 $B^-\to \ov D'_{s1}(2536)^-D^0$ & $1.3\times 10^{-4}$ & $3.4\times 10^{-3}$ & & \\
 $B^-\to \ov D^*_{s2}(2572)^-D^0$ & -- & 0 & & \\
 \hline
 $B^-\to \ov D^*_{s0}(2317)^-D^{*0}$ & $5.0\times 10^{-4}$ & & & \\
 $B^-\to \ov D_{s1}(2460)^-D^{*0}$ & $1.2\times 10^{-2}$ & & & \\
 $B^-\to \ov D'_{s1}(2536)^-D^{*0}$ & $5.3\times 10^{-4}$ & & & \\
 $B^-\to \ov D^*_{s2}(2572)^-D^{*0}$ & -- &  & & \\
\end{tabular}
\end{center}
\end{table}

\begin{table}[p]
\caption{Same as Table \ref{tab:chargedB} except for neutral $B$
and $B_s$ mesons.
 }  \label{tab:neutralB}
\begin{center}
\begin{tabular}{l c c c c l  }
Decay & This Work & KV \cite{Katoch} &  CM \cite{Castro} & KLO \cite{Kim}& Expt \\
\hline
 $\ov B^0\to D^*_0(2308)^+\pi^-$ & $2.6\times 10^{-4}$ & $4.1\times 10^{-4}$ & & &  \\
 $\ov B^0\to D_1(2427)^+\pi^-$ & $6.8\times 10^{-4}$ & $1.2\times 10^{-4}$ & & &  \\
 $\ov B^0\to D'_1(2420)^+\pi^-$ & $1.0\times 10^{-3}$ & $2.4\times 10^{-3}$ & & &  \\
 $\ov B^0\to D^*_2(2460)^+\pi^-$ & $6.1\times 10^{-4}$ & $7.1\times 10^{-5}$
 & $4.1\times 10^{-4}$ & $3.3\times 10^{-4}$ & $<2.2\times 10^{-3}$\cite{PDG}  \\
 \hline
 $\ov B^0\to D^*_0(2308)^+\rho^-$ & $6.4\times 10^{-4}$ &  \\
 $\ov B^0\to D_1(2427)^+\rho^-$ & $1.6\times 10^{-3}$ & \\
 $\ov B^0\to D'_1(2420)^+\rho^-$ & $2.6\times 10^{-3}$ & & & &  \\
 $\ov B^0\to D^*_2(2460)^+\rho^-$ & $1.7\times 10^{-3}$ & & $1.1\times 10^{-3}$
 & $9.2\times 10^{-4}$ & $<4.9\times 10^{-3}$\cite{PDG} \\
 \hline
 $\ov B^0\to D^*_0(2308)^+\bar D_s^-$ & $7.3\times 10^{-4}$ & $2.6\times 10^{-3}$ & &\\
 $\ov B^0\to D_1(2427)^+\bar D_s^-$ & $8.8\times 10^{-4}$ & $1.3\times 10^{-3}$ & & \\
 $\ov B^0\to D'_1(2420)^+\bar D_s^-$ & $1.2\times 10^{-3}$ & $4.9\times 10^{-3}$ & \\
 $\ov B^0\to D^*_2(2460)^+\bar D_s^-$ & $3.8\times 10^{-4}$ & $1.0\times 10^{-4}$
 & $2.7\times 10^{-4}$ & $4.6\times 10^{-4}$ &  \\
 \hline
 $\ov B^0\to D^*_0(2308)^+\bar D_s^{*-}$ & $3.2\times 10^{-4}$ & & &\\
 $\ov B^0\to D_1(2427)^+\bar D_s^{*-}$ & $5.5\times 10^{-4}$ & & & \\
 $\ov B^0\to D'_1(2420)^+\bar D_s^{*-}$ & $1.5\times 10^{-3}$ & & \\
 $\ov B^0\to D^*_2(2460)^+\bar D_s^{*-}$ & $1.0\times 10^{-3}$
 & $1.0\times 10^{-4}$ & $1.1\times 10^{-3}$ & $1.1\times 10^{-3}$ \\
 \hline
 $\ov B^0\to \ov D^*_{s0}(2317)^-D^+$ & ($8.6\times 10^{-4}$) & 0 & & & see text \\
 $\ov B^0\to \ov D_{s1}(2460)^-D^+$ &  ($2.8\times 10^{-3}$) & $3.4\times 10^{-3}$ & & & see  text \\
 $\ov B^0\to \ov D'_{s1}(2536)^-D^+$ & $1.2\times 10^{-4}$ & $3.3\times 10^{-3}$ & & \\
 $\ov B^0\to \ov D^*_{s2}(2572)^-D^+$ & -- & 0 & & \\
 \hline
 $\ov B^0\to \ov D^*_{s0}(2317)^-D^{*+}$ & $4.6\times 10^{-4}$ & & & \\
 $\ov B^0\to \ov D_{s1}(2460)^-D^{*+}$ & $1.1\times 10^{-2}$ & & & \\
 $\ov B^0\to \ov D'_{s1}(2536)^-D^{*+}$ & $4.9\times 10^{-4}$ & & & \\
 $\ov B^0\to \ov D^*_{s2}(2572)^-D^{*+}$ & -- &  & & \\
 \hline
 $\ov B_s^0\to D^*_{s0}(2317)^+\pi^-$ & $3.3\times 10^{-4}$ & & \\
 $\ov B_s^0\to D_{s1}(2460)^+\pi^-$ & $5.2\times 10^{-4}$ & & \\
 $\ov B_s^0\to D'_{s1}(2536)^+\pi^-$ & $1.5\times 10^{-3}$ & & \\
 $\ov B_s^0\to D^*_{s2}(2572)^+\pi^-$ & $7.1\times 10^{-4}$ & & \\
 \hline
 $\ov B_s^0\to D^*_{s0}(2317)^+\rho^-$ & $8.3\times 10^{-4}$ & & \\
 $\ov B_s^0\to D_{s1}(2460)^+\rho^-$ & $1.3\times 10^{-3}$ & & \\
 $\ov B_s^0\to D'_{s1}(2536)^+\rho^-$ & $3.8\times 10^{-3}$ & & \\
 $\ov B_s^0\to D^*_{s2}(2572)^+\rho^-$ & $1.9\times 10^{-3}$ & & \\
\end{tabular}
\end{center}
\end{table}

\subsection{Results and discussions}
Given the decay constants and form factors discussed in Sec. III,
we are ready to study the $B$ decays into $p$-wave charmed mesons.
The predicted branching ratios are shown in Tables
\ref{tab:chargedB} and \ref{tab:neutralB}.  The experimental
results are taken from PDG \cite{PDG} and Belle \cite{BelleD}. For
$B^-\to D_2^{*0}\pi^-$ we combine the Belle measurements
\cite{BelleD}
 \be
 \B(B^-\to D_2^{*0}\pi^-)\B(D_2^{*0}\to D^+\pi^-) &=& (3.4\pm0.3\pm0.6\pm0.4)\times
 10^{-4},
 \non \\
 \B(B^-\to D_2^{*0}\pi^-)\B(D_2^{*0}\to D^{*+}\pi^-) &=& (1.8\pm0.3\pm0.3\pm0.2)\times
 10^{-4},
 \en
to arrive at
 \be
 \B(B^-\to D_2^{*0}\pi^-)\B(D_2^{*0}\to D^+\pi^-,D^{*+}\pi^-)=(5.5\pm
 0.8)\times 10^{-4}.
 \en
Using $\B(D_2^{*0}\to D^+\pi^-,D^{*+}\pi^-)=2/3$ followed from the
assumption that the $D_2^{*0}$ width is saturated by $D\pi$ and
$D^*\pi$, we are led to $\B(B^-\to D_2^{*0}\pi^-)=(7.8\pm
1.4)\times 10^{-4}$ as shown in Table \ref{tab:chargedB}.

From Table \ref{tab:chargedB} we see that except
$D_1(2427)^0\pi^-$ the predictions of $\B(B^-\to D^{**}\pi^-)$
agree with experiment. It is worth mentioning that the ratio
 \be
 R={\B(B^-\to D_2^*(2460)^0\pi^-)\over \B(B^-\to D'_1(2420)^0\pi^-)}
 \en
is measured to be $0.77\pm0.15$ by Belle \cite{BelleD} and
$1.8\pm0.8$ by CLEO \cite{Galik}. The early prediction by Neubert
\cite{Neubert} yields a value of 0.35. Our prediction of $R=0.61$
is in accordance with the data. However, the predicted rate for
$D_1(2427)^0\pi^-$ is too small by a factor of 2. This is ascribed
to a destructive interference between color-allowed and
color-suppressed tree amplitudes because the form factors
$V_0^{BD_1^{1/2}}$ and $V_0^{BD_1^{3/2}}$ have signs opposite to
that of $F_1^{B\pi}$ as required by heavy quark symmetry (see Eq.
(\ref{BtoDpi}) and Table \ref{tab:BtoD1a}). In contrast, the
production of $D_1(2427)^+\pi^-$ is larger than $D_1(2427)^0\pi^-$
by a factor of about 2 because the former does not receive the
destructive contribution from internal $W$-emission. Hence, a
measurement of the ratio $D_1(2427)^0\pi^-/D_1(2427)^+\pi^-$ can
be used to test the relative signs of various form factors as
implied by heavy quark symmetry. Note that for the
$D_1^{1/2}-D_1^{3/2}$ mixing angle we use $\theta=17^\circ$ [see
Eq. (\ref{Dmixing})]. If a negative value of $-17^\circ$ is
employed, the decay $B^-\to D_1(2427)^0\pi^-$ will be severely
suppressed with a branching ratio of order $6\times 10^{-6}$. This
means that the $D_1^{1/2}-D_1^{3/2}$ mixing angle is preferred to
be positive. In their study Katoch and Verma \cite{Katoch}
obtained a small branching ratio for $B^-\to D_0^*(2308)^0\pi^-$
as they assumed a vanishing decay constant for $D_0^*$. As
stressed before, this decay constant is comparable to $f_D$
because of charm and light quark mass imbalance and a rough
estimate yields $f_{D_0^*}\approx 160$ MeV [cf. Eq. (\ref{fD0})].
Consequently, the contribution from internal $W$-emission will
account for the aforementioned discrepancy between theory and
experiment. Moreover, the ratio $D_0^{*+}\pi^-/D_0^{*0}\pi^-$ is
predicted to be 0.34 instead of unity because of the absence of
the color-suppressed tree contribution to the former.

At a first glance, it appears that the prediction $\B(B^-\to
D'_1(2420)^0\rho^-)=2.8\times 10^{-3}$ already exceeds the
experimental limit $1.4\times 10^{-3}$ \cite{PDG}. However, it
should be noticed that the ${D'_1}^0\rho^-$ rate is about three
times larger than that of ${D'_1}^0\pi^-$ as expected from the
factorization approach and from the ratio $f_\rho/f_\pi\approx
1.6$ (see Table \ref{tab:ratios} below). Hence, it appears that
the present limit on $D_1'\rho^-$ is not consistent with the
observed rate of $D_1'\pi^-$. Of course, it is crucial to measure
$B\to D'_1(2420)\rho$ in order to clarify the issue.

Apart from the external $W$-emission diagram, the
$D^{**}\pi(\rho)$ productions in neutral $B$ decays also receive
$W$-exchange contributions which are neglected in the present
work. This will constitute a main theoretical uncertainty for $\ov
B^0\to D^{**+}\pi^-(\rho^-)$.

For $B\to \ov DD_s^{**}$ decays, their factorizable amplitudes are
rather simple
 \be
 A(\ov B\to D\ov D_s^{**})={G_F\over\sqrt{2}}V_{cb}V_{cs}^*\,a_1\la
 \ov D_s^{**}|(\bar sc)|0\ra\la D|(\bar c b)|\ov B\ra.
 \en
The decay rates are thus governed by the decay constants. Since
the tensor meson cannot be produced from the $V-A$ current, the
$B$ decay into $\ov D_{s2}^*D^{(*)}$ is prohibited under the
factorization hypothesis. For $D_{s0}^*$ and $D_{s1}$ mesons, it
is expected that $f_{D_{s0}^*}\sim f_{D_{s1}(2460)}\gg
f_{D'_{s1}(2536)}$ as $f_{D_{s1}^{1/2}}=f_{D_{s0}^*}$ and
$f_{D_{s1}^{1/2}}=0$ in the heavy quark limit [see Eq.
(\ref{fHQS})]. When the charmed quark mass is finite, the
$D_s^{1/2}\!-\!D_s^{3/2}$ mixing effect has to be taken into
account via
  \be
 f_{D_{s1}} &=&
 f_{D_{s1}^{1/2}}\cos\theta_s+f_{D_{s1}^{3/2}}\sin\theta_s, \non
 \\
 f_{D'_{s1}} &=&
 -f_{D_{s1}^{1/2}}\sin\theta_s+f_{D_{s1}^{3/2}}\cos\theta_s,
 \en
where use of Eqs. (\ref{Dsmixing}) and (\ref{decayconst}) has been
made. Consequently, it is anticipated that $\ov DD_{s0}^*\gsim \ov
DD_{s1}(2460)\gg \ov DD_{s1}(2536)$ after taking into account
phase space corrections. However, the recent data on $\ov
DD_s^{**}$ production indicate a surprise.

The results of Belle measurements read \cite{FPCP03}
 \be
 \B[B\to \ov DD_{s0}^*(2317)]\B[D_{s0}^*(2317)\to D_s\pi^0] &=&
 (8.5^{+2.1}_{-1.9}\pm2.6)\times 10^{-4}, \non \\
 \B[B\to \ov DD_{s1}(2460)]\B[D_{s1}(2460)\to D_s^*\pi^0] &=&
 (17.8^{+4.5}_{-3.9}\pm5.3)\times 10^{-4}, \non \\
 \B[B\to \ov DD_{s1}(2460)]\B[D_{s1}(2460)\to D_s\gamma] &=&
 (6.7^{+1.3}_{-1.2}\pm2.0)\times 10^{-4}.
 \en
Since  $D_{s0}^*(2317)$ is dominated by its hadronic decay to
$D_s\pi^0$,\footnote{The upper limit on the ratio
$\Gamma(D_{s0}^*\to D_s^*\gamma)/\Gamma(D_{s0}^*\to
D_s\pi^0)<0.059$ was set recently by CLEO \cite{CLEO}.} the
branching ratio of $B\to DD_{s0}^*(2317)$ is of order $0.9\times
10^{-3}$. This means that the production rate of $\ov D^0D_{s0}^*$
in $B$ decays is smaller than $\B(B^+\to \ov
D^0D_s)=(1.3\pm0.4)\%$ \cite{PDG} by one order of magnitude. Since
this decay proceeds only via external $W$-emission, it can be used
to determine the decay constant of $D_{s0}^*$. It is found that
$f_{D_{s0}^*}\sim 60$ MeV \footnote{Interestingly, this happens to
be the original estimate of $f_{D_{s0}^*}$ made in \cite{CH} for
$D_{s0}^*$ in the four-quark state.} which seems too small,
recalling that $f_{D_0^*}\sim 160$ MeV is needed to account for
the production of $D_0^*\pi^-$. To estimate the branching ratios
of $D_s^*\pi^0$ and $D_s\gamma$ in $D_{s1}(2460)$ decay, we need
some experimental and theoretical inputs. Besides the values
$0.38\pm0.12$ and $0.63\pm0.21$ measured by Belle for the ratio
$D_s\gamma/D_s^*\pi^0$ [see Eq. (\ref{gammatopi0})], the ratios
$D_s^*\gamma/D_s^*\pi^0$ and $D_s\pi^+\pi^-/D_s^*\pi^0$ are found
to be less than 0.16 and 0.08, respectively, by CLEO \cite{CLEO}.
Theoretically, the $M1$ transition $D_{s1}\to D_{s0}^*\gamma$
turns out to be quite small \cite{Godfrey}. Assuming that the
$D_{s1}(2460)$ decay is saturated by
$D_s^*\pi^0,~D_s\gamma,~D_s^*\gamma$ and $D_s\pi\pi$, we are led
to
 \be
 0.56\lsim \B(D_{s1}(2460)\to D_s^*\pi^0)\lsim 0.75,
 \en
where the averaged value $0.44\pm0.10$ for the ratio
$D_s\gamma/D_s^*\pi^0$ has been employed. This in turn implies
$\B[B\to \ov DD_{s1}(2460)]=(1.5\sim 4.4)\times 10^{-3}$. As a
result, the decay constant of $D_{s1}(2460)$ is found to be
 \be
 f_{D_{s1}(2460)}\approx 100-175\,{\rm MeV}.
 \en
This is in sharp contrast to the heavy quark symmetry expectation
that the decay constants of $D_{s0}^*$ and $D_{s1}$ become
identical in the heavy quark limit. Therefore, the large disparity
between $f_{D_{s0}^*}$ and $f_{D_{s1}}$ is a surprise. It is not
clear to us what is the underlying reason for the discrepancy
between theory and experiment.

Although the $B$ decay into $\ov D_{s2}^*D^{(*)}$ is prohibited
under the factorization hypothesis, nevertheless it can be induced
via final-state interactions (FSIs) and/or nonfactorizable
contributions. For example, it can be generated via the
color-allowed decay $B^-\to \ov D_sD_2^{*0}$ followed by the
rescattering process $\ov D_sD_2^{*0}\to \ov D_{s2}^*D^0$. Since
the nonfactorizable term is of order $c_2/N_c$ with the Wilson
coefficient $c_2(m_b)\approx -0.20$, it is likely to be suppressed
relative to the FSIs. Hence, an observation of $B\to\ov
D_{s2}^*D^{(*)}$ could imply the importance of final-state
rescattering effects.

Since heavy quark symmetry implies $f_{D'_{s1}(2536)}\ll
f_{D_{s1}(2460)}$, it is important to measure the $B$ decay into
$\ov D'_{s1}(2536)D^{(*)}$ to see if it is suppressed relative to
$\ov D_{s1}(2460)D^{(*)}$ to test heavy quark symmetry. In order
to compute $\ov B\to \ov D_s^{**}D^*$ we have fixed the decay
constants of $D_{s0}^*$, $D_{s1}^{1/2}$ and $D_{s1}^{3/2}$ as
shown in Eq. (\ref{decayconst}) so that the resulting production
rates of $\ov D_{s0}^*D^0$ and $\ov D_{s1}D^0$ are consistent with
experiment (see Tables \ref{tab:chargedB} and \ref{tab:neutralB}).
We see that the decay $\ov B\to\ov D_{s1}(2460)D^{*0}$ has the
largest branching ratio of order $10^{-2}$ in two-body hadronic
$B$ decays involving a $p$-wave charmed meson in the final state.
It is essential to test all these anticipations in the near
future. A recent CLEO measurement yields \cite{CLEODsD}
 \be
 \B[B^-\to (\bar D_s+\bar D_s^*)(D_1^0+{D'_1}^0+D_2^{*0})]=(2.73\pm
 0.78\pm0.48\pm0.68)\%.
 \en
Our prediction of $6\times 10^{-3}$ is slightly small.

It is interesting to consider the ratios $\Gamma(\ov B\to
D^{**}V)/\Gamma(\ov B\to D^{**}P)$ and $\Gamma(\ov B\to
D_s^{**}V)/\Gamma(\ov B\to D_s^{**}P)$. The calculated results are
shown in Table \ref{tab:ratios}. Several remarks are in order: (i)
The ratios $D^{**0}\rho^-/D^{**0}\pi^-$ and
$D^{**+}\rho^-/D^{**+}\pi^-$ for $D^{**}=D_0^*$ and $D_1$ are not
the same as the former receives an additional color-suppressed
internal $W$-emission contribution. (ii) Whether the ratio
$D^{**}V/D^{**}P$ is greater than unity or not depends essentially
on the ratio of the decay constants. For example, $D_0^*\bar
D_s^*/D_0^*\bar D_s=0.43$ if the decay constants of $D_s^*$ and
$D_s$ are similar, while $D_0^{*+}\rho^- /D_0^{*+}\pi^-=2.5$ for
$f_\rho/f_\pi=1.6$\,. It should be stressed that the proximity of
the ratio $D^{**+}\rho^- /D^{**+}\pi^-$ to 2.5 has less to do with
the three degrees of freedom of $\rho$, rather it is mainly
related to the decay constant ratio of $f_\rho/f_\pi$. (iii) The
ratio of $D_s^{**}\rho^-/D_s^{**}\pi^-$ in $B_s$ decay is the same
as $D^{**+}\rho^-/D^{**+}\pi^-$ in $B$ decay as they proceed via
external $W$-emission.

\begin{table}[ht]
\caption{The ratios $\Gamma(\ov B\to D^{**}V)/\Gamma(\ov B\to
D^{**}P)$ and $\Gamma(\ov B\to D_s^{**}V)/\Gamma(\ov B\to
D_s^{**}P)$. The last column is for the ratio $\Gamma(\ov B_s\to
D_s^{**}\rho^-)/\Gamma(\ov B_s\to D_s^{**}\pi^-)$.
 } \label{tab:ratios}
\begin{center}
\begin{tabular}{c c c c c c}
  $D^{**}(D_s^{**})$ & ${D^{**0}\rho^-\over D^{**0}\pi^-}$ & ${D^{**+}\rho^-
  \over D^{**+}\pi^-}$  &  ${D^{**}\bar D_s^*\over D^{**}\bar D_s}$ &
  ${\bar D_s^{**}D^*\over \bar D_s^{**}D}$ &
  ${D_s^{**}\rho^-\over D_s^{**}\pi^-}$ \\
 \hline
 $D_0^*(D_{s0}^*)$ & 1.7 & 2.4  & 0.43 & 0.54 & 2.5 \\
 $D_1(D_{s1})$ & 3.1 & 2.4  & 0.63 & 3.8 & 2.5 \\
 $D_1'(D_{s1}')$ & 2.6 & 2.6 & 1.2 & 4.0 & 2.6 \\
 $D_2^*(D_{s2}^*)$ & 2.7 & 2.7 & 2.7 & -- & 2.7 \\
\end{tabular}
\end{center}
\end{table}

Because the scalar resonances $D_0^*$ and $D_1$ have widths of
order 300 MeV, we have checked the finite width effects on their
production in $B$ decays and found that the conventional narrow
width approximation is accurate enough to describe the production
of broad resonances owing to the large energy released in hadronic
two-body decays of $B$ mesons.

\subsection{Comparison with other works}
The decays $B\to D^{**}(\pi,D_s)$ and $B\to D_s^{**}D$  have been
studied previously by Katoch and Verma (KV) \cite{Katoch}. L\'opez
Castro and Mu\~noz (CM)\cite{Castro}, Kim, Lee and Oh (KLO)
\cite{Kim}, and Lee \cite{Lee} also have a similar study with
focus on the tensor charmed meson production. We shall comment on
their works separately.

In the paper of KLO, the $B\to D_2^*$ form factors are evaluated
using the ISGW2 model. However, the predicted rates for $D_2^*\pi$
and $D_2^*\rho$ by KLO are smaller than ours by a factor 2 (see
Tables \ref{tab:chargedB} and \ref{tab:neutralB}); that is, their
$B\to D_2^*$ form factor differs from ours roughly by a factor of
$\sqrt{2}$. In contrast, the results for $D_2^* D_s^{(*)}$ are
similar owing to the fact that KLO employ $f_{D_s}=280$ MeV and
$f_{D_s^*}=270$ MeV which are larger than ours [see Eq.
(\ref{decayconst})]. Instead of employing the ISGW2 model for
$B\to D_2$ form factors, Lee applied the QCD sum rule to study the
universal function $\tau_{3/2}(\omega)$ to the leading order
\cite{Lee}. He found $\B(B^-\to D_2^{*0}\pi^-)=12.9\times 10^{-4}$
for $a_1=1.15$, which is two times larger than ours.

The work of KV and CM is based on the original version of the ISGW
model. However, as stressed by KLO \cite{Kim}, the exponentially
decreasing behavior of the form factors in the ISGW model is not
realistic and justified. This has been improved in the ISGW2 model
which provides a more realistic description of the form-factor
behavior at large $(q^2_m-q^2)$. The values of form factors at
small $q^2$ in the ISGW2 model can be a few times larger than that
obtained in the ISGW model as the maximum momentum transfer
$q^2_m$ in $B$ decays is large. However, the expected form-factor
suppression does not appear in the calculations of KV as they
calculated the form factors at $q_m^2$ and then employed them even
at low $q^2$. In contrast, CM did compute the form factors at
proper $q^2$. The fact that CM and KLO have similar results for
$B\to D_2^*\pi(\rho)$ and $B\to D_2^*\bar D_s^{(*)}$ (see Tables
\ref{tab:chargedB} and \ref{tab:neutralB}) is surprising as the
$B\to D_2^*$ transition is evaluated in two different versions of
the ISGW model. To check this, we find that $\eta(q^2)\equiv
k(q^2)+b_+(q^2)(m_B^2-m^2_{D_2})+b_-(q^2)q^2=0.286$ and 0.386 for
$q^2=m_\pi^2$ in the ISGW and ISGW2 models, respectively, which in
turn imply the respective branching ratios, $6.7\times 10^{-4}$
and $3.8\times 10^{-4}$ for $B^-\to D_2^{*0}\pi^-$. Therefore, our
result obtained in the ISGW model is consistent with that of CM
and the estimate of $\eta(q^2)$ within the ISGW2 model by KLO is
likely too small, as noted in passing.

The axial-vector charmed meson production considered by KV is for
$D_1(^1P_1)$ and $D_1(^3P_1)$ rather than $D_1^{1/2}$ and
$D_1^{3/2}$. Therefore, the expressions of the decay amplitudes
involving $D_1$ or $D_{s1}$ by KV should be modified by taking
into account a proper wave function combination, Eq. (\ref{PjJ}).
For the decay constants, KV assumed that $f_{D_0^*}=0$ and
$f_{D_1(^1P_1)}=0$. As a consequence, $B^-\to D_0^{*0}\pi^-$ is
too small compared to experiment and the decay into $\ov
D_{s0}^*D$ is not allowed. This is not consistent with the heavy
quark symmetry relation $f_{D_{s1}^{1/2}}=f_{D^*_{s0}}$. Finally,
the predicted rate of $B^-\to D_2^{*0}\pi^-$ by KV is too small by
one order of magnitude compared to experiment. This is ascribed to
a missing factor of $(m_B/m_T)^2$ in their calculation of decay
rates.

\section{Conclusions}
The hadronic decays of $B$ mesons to a $p$-wave charmed meson in
the final state are studied. Specifically we focus on the
Cabibbo-allowed decays $\ov B\to D^{**}\pi(\rho),~D^{**}\bar
D_s^{(*)},~\bar D^{**}_sD^{(*)}$ and $\ov B_s\to
D_s^{**}\pi(\rho)$. The main conclusions are as follows:

\begin{itemize}
 \item
 We apply heavy quark effective theory in which heavy quark
symmetry and chiral symmetry are unified to study the strong
decays of $p$-wave charmed mesons and determine the magnitude of
the $D_1^{1/2}\!-\!D_1^{3/2}$ mixing angle. In contrast, the
present upper limits on the widths of $D_{s1}(2460)$ and
$D'_{s1}(2536)$ do not provide any  constraints on the
$D_{s1}^{1/2}\!-\!D_{s1}^{3/2}$ mixing angle $\theta_s$.
Therefore, we appeal to the quark potential model to extract
$\theta_s$.

 \item
 Various form factors for $B\to D^{**}$ transitions and their $q^2$
dependence are studied using the improved version of the
Isgur-Scora-Grinstein-Wise quark model. Heavy quark symmetry
constraints are respected in this model calculation.
 \item
 The predicted branching ratios for $B^-\to D^{**}\pi^-$  agree with
experiment except $D_1(2427)^0\pi^-$. The
$D_1^{1/2}\!-\!D_1^{3/2}$ mixing angle is preferred to be positive
in order to avoid a severe suppression on the production of
$D_1(2427)^0\pi^-$. The decay $B^-\to D'_1(2420)^0\rho^-$ is
predicted at the level of $3\times 10^{-3}$. Although it exceeds
the present experimental limit of $1.4\times 10^{-3}$, it leads to
the ratio $D'_1(2420)\rho^-/D'_1(2420)\pi^-\approx 2.6$ as
expected from the factorization approach and from the ratio
$f_\rho/f_\pi\approx 1.6$\,. Therefore, it is crucial to have a
measurement of this mode to test the factorization hypothesis.
 \item
 The predicted rate for $B^-\to D_1(2427)^0\pi^-$ is too small by a
factor of 2 owing to a destructive interference between
color-allowed and color-suppressed tree amplitudes as the relevant
form factors for $B\to D_1^{1/2}$ and $B\to D_1^{3/2}$ transitions
are negative. It is crucial to measure the production of
$D_1(2427)^+\pi^-$ to see if it is larger than $D_1(2427)^0\pi^-$
by a factor of about 2 because the former does not receive the
internal $W$-emission contribution.
 \item
 Under the factorization hypothesis, the production of $\ov
D_{s2}^*D^{(*)}$ in $B$ decays is prohibited as the tensor meson
cannot be produced from the $V-A$ current. Nevertheless, the
decays $B\to\ov D_{s2}^*D^{(*)}$ can be induced via final-state
interactions and/or the nonfactorizable contributions.  Since the
latter is suppressed by the order of $c_2/N_c$, an observation of
$B\to\ov D_{s2}^*D^{(*)}$ could imply the importance of
final-state rescattering effects.
 \item
 For $\ov B\to \ov D_s^{**}D$ decays, it is expected that $\ov
D_{s0}^*D\gsim \ov D_{s1}D$ as the decay constants of the
multiplet $(D_{s0}^*,D_{s1})$ become identical in the heavy quark
limit. The experimental observation indicates that $\ov
D_{s0}^*D/\ov D_{s1}D\approx (0.2-0.6)$ and that
$f_{D_{s1}}\approx (100-175)$ MeV and $f_{D_{s0}^*}\sim 60$ MeV.
The less abundant production of $\ov D_{s0}^*D$ than $\ov D_{s1}D$
and the large disparity between $f_{D_{s1}}$ and $f_{D_{s0}^*}$
are surprising. The reason for the discrepancy between theory and
experiment remains a puzzle. In the meantime, it is also important
to measure the decay to $\ov D'_{s1}(2536)D^{(*)}$ to see if it is
suppressed relative to $\ov D_{s1}(2460)D^{(*)}$ to test the heavy
quark symmetry relation $f_{D_{s1}(2536)}\ll f_{D_{s1}(2460)}$.

\end{itemize}

\vskip 2.5cm \acknowledgments We are grateful to Chuang-Hung Chen
and Taekoon Lee for valuable discussions. This work was supported
in part by the National Science Council of R.O.C. under Grant No.
NSC91-2112-M-001-038.

\vskip 2cm \noindent  {\it Note Added:}

\vskip0.2cm \noindent After this work was completed, we noticed
the appearance of the related works on $B\to \ov DD_s^{**}$ decays
by
C.H. Chen and H.n. Li, hep-ph/0307075;
A. Datta and P.J. O'Donnell, hep-ph/0307106;
M. Suzuki, hep-ph/0307118.
%


\end{document}